%% file: main.tex
\documentclass[a4paper,numberwithinsect,USenglish]{lipics-v2021}

\title{On~the~Descriptive~Complexity~of Vertex~Deletion~Problems}

\titlerunning{Descriptive Complexity of Vertex Deletion Problems}

\author{Max Bannach}{European Space Agency, Advanced Concepts Team, Noordwijk,
  The Netherlands}{max.bannach@esa.int}{https://orcid.org/0000-0002-6475-5512}{}
\author{Florian Chudigiewitsch}{Universität zu Lübeck, Germany}{fch@tcs.uni-luebeck.de}{https://orcid.org/
0000-0003-3237-1650}{}
\author{Till Tantau}{Universität zu Lübeck, Germany}{tantau@tcs.uni-luebeck.de}{}{}

\authorrunning{M. Bannach and F. Chudigiewitsch and T. Tantau}
\Copyright{M. Bannach and F. Chudigiewitsch and T. Tantau} 


\category{} 

\relatedversion{} 

\nolinenumbers 

\hideLIPIcs  

\EventEditors{John Q. Open and Joan R. Access}
\EventNoEds{2}
\EventLongTitle{42nd Conference on Very Important Topics (CVIT 2016)}
\EventShortTitle{CVIT 2016}
\EventAcronym{CVIT}
\EventYear{2016}
\EventDate{December 24--27, 2016}
\EventLocation{Little Whinging, United Kingdom}
\EventLogo{}
\SeriesVolume{42}
\ArticleNo{23}

\theoremstyle{plain}
\newtheorem{problem}[theorem]{Problem}
\newtheorem{fact}[theorem]{Fact}

\usepackage{listings}
\lstdefinelanguage{pseudocode}{
  morekeywords={
    algorithm,method,new,and,not,
    if,then,else,while,do,repeat,until,seq,
    seqdo,return,call,
    for,pardo,foreach,print,output,input,exit,
    break,loop,end,begin,goto,par,global,local,
    read,write,stop,idle,procedure,function,
    throw,catch
  },
  sensitive=true,
  morecomment=[l]{//},
  morestring=[b]",
  morestring=[s]{``}{''},
}
\lstdefinestyle{pseudocode}{
  language=pseudocode,
  basicstyle=\small\rmfamily,
  commentstyle=\upshape\color{black!50},
  keywordstyle=\bfseries\itshape,
  identifierstyle=\itshape,
  stringstyle=\rmfamily,
  columns=fullflexible,
  mathescape,
  literate={<-}{{$\gets$\ }}2,
  numbers=left,
  numberstyle=\scriptsize\sffamily,
}

\usepackage{booktabs}

\usepackage{mathtools}

\DeclareMathOperator{\adj}{\sim}
\DeclareMathOperator{\nadj}{\not\sim}

\newcommand\Para{\mathrm{para\text-}}

\newcommand\Class[1]{%
  \mathchoice%
  {\text{\normalfont\small$\mathrm{#1}$}}%
  {\text{\normalfont\small$\mathrm{#1}$}}%
  {\text{\normalfont$\mathrm{#1}$}}%
  {\text{\normalfont$\mathrm{#1}$}}%
}
\newcommand\PClass{\mathrm{p\text{\normalfont-}}\Class}

\newcommand\PLangVD[1]{\PLang{vd}_{\mathrm{#1}}}
\newcommand\PVD[1]{\PClass{VD}_{\mathrm{#1}}}

\newenvironment{parameterizedproblem}%
{%
  \leavevmode\nobreak\par
  \begin{list}%
    {}%
    {%
      \def\labelstyle{\itshape}
      \setlength{\topsep}{0pt}%
      \settowidth{\labelwidth}{\labelstyle Parameter:}%
      \setlength{\leftmargin}{\labelwidth}%
      \addtolength{\leftmargin}{\labelsep}%
      \setlength{\itemsep}{0pt}%
      \setlength{\parsep}{0pt}%
    }%
      \def\instance{\item[\labelstyle Instance:]}%
      \def\parameter{\item[\labelstyle Parameter:]}%
      \def\question{\item[\labelstyle Question:]}%
    }%
    {%
  \end{list}%
}

\newcommand{\Lang}[1]{\text{\normalfont\textsc{#1}}}
\newcommand{\PLang}[2][]{\mathrm{p}_{#1}\Lang{-#2}}

\newcommand\coloneq{\mathrel{\raise.4pt\hbox{:}{=}}}
\newcommand\eqcolon{\mathrel{{=}\raise.4pt\hbox{:}}}

\usepackage[]{tcs-automove}

\usepackage{xfrac}
\usepackage{tikz}
\usetikzlibrary{graphs, arrows.meta, positioning, backgrounds, shapes, decorations.pathmorphing,}

\definecolor{ba.yellow}{RGB}{252,190,18}
\definecolor{ba.gray}{RGB}{153,153,156}
\colorlet{ba.blue}{blue!80!black}
\colorlet{ba.red}{red!90!black}
\definecolor{ba.orange}{RGB}{233,116,81}
\definecolor{ba.pine}{RGB}{67,154,134}
\colorlet{ba.green}{green!50!black}
\definecolor{ba.violet}{RGB}{88, 53, 94}

\def\AtMost{\textcolor{black!50}{p \rlap{\,$\preceq$}}}
\def\AtLeast{\textcolor{black!50}{\llap{$\preceq$\,} p\rlap.}}
\def\Between{\textcolor{black!50}{\llap{$\preceq$\,} p\rlap{\,$\preceq$}}}

\usepackage{tikz}
\usetikzlibrary{graphs, arrows.meta, positioning, backgrounds, shapes, decorations.pathmorphing,}
\tikzset{
  every picture/.style={semithick},%
  > = {Stealth[round,sep]},
  grayed/.style = { black!30 },
  node/.style={
    draw,
    circle,
    minimum size=5mm,
    inner sep=0.5pt,
    font=\footnotesize,
    fill=white
  },%
  vertex/.style={ 
    node 
  },
  small node/.style={
    node,minimum size=4.5pt,
    inner sep=0pt,
    outer sep=0pt,
    font=\tiny
  },%
  on/.style={fill=white,inner sep=-.4pt,circle},
  mapsto/.style={|[sep]->,blue!50!black},
  arbitrary/.style={dashed,draw=black!50},
  present/.style={black},
  missing/.style={black!10},
  virtual node/.style={node,draw=none,fill=none},
  white node/.style={node},
  black node/.style={node, fill=black, text=white},
  gray node/.style={node, arbitrary, fill=black!20},
  red node/.style={node, red!60!black, text=white},
  blue node/.style={node, blue!70!black, text=white},
  green node/.style={node, green!50!black, text=white},
  hilight/.style={red!80!black},
  new/.style={
    line width=.4pt,
    double distance=.8pt,
    draw=white,
    double=lipicsYellow,
  },
  new edge/.style={
    draw=lipicsYellow,
    thick
  }
}

\begin{document}

\maketitle

\begin{abstract}
  Vertex deletion problems for graphs are studied intensely in
  classical and parameterized complexity theory. They ask whether we
  can delete at most $k$ vertices from an input graph such that the
  resulting graph has a certain property. Regarding $k$ as the
  parameter, a dichotomy was recently shown based on the number of
  quantifier alternations of first-order formulas that describe the
  property. In this paper, we refine this classification by moving 
  from quantifier alternations to individual quantifier patterns and
  from a dichotomy to a trichotomy, resulting in a complete classification of
  the complexity of vertex deletion problems based on their quantifier
  pattern. The more fine-grained approach uncovers new tractable
  fragments, which we show to not only lie in $\Class{FPT}$, but even
  in parameterized constant-depth circuit complexity classes. On the other
  hand, we show that vertex deletion becomes intractable already for
  just one quantifier per alternation, that is, there is a formula of
  the form $\forall x\exists y\forall z (\psi)$, with $\psi$
  quantifier-free, for which the vertex deletion problem is $\Class
  W[1]$-hard. The fine-grained analysis also allows us to uncover
  differences in the complexity landscape when we consider different
  kinds of graphs and more general structures: While basic graphs
  (undirected graphs without self-loops), undirected graphs, and
  directed graphs each have a different frontier of tractability,
  the frontier for arbitrary logical structures coincides with that of
  directed graphs.
\end{abstract}

\section{Introduction}

A recent research topic in parametrized complexity are \emph{distance
to triviality problems.} We are asked how many modification steps (the
``distance'') we need to apply to a logical structure in order to 
transform it into a ``trivial'' one -- which can mean anything from ``no
edges at all'' to ``no cycles'' or even more exotic properties like ``no cycles of odd length.'' Such
problems have been found highly useful in modern algorithm
design~\cite{Abu-KhzamCFLSS04,AgrawalR22,Dumas023,GuoHN04} and are now an important test bed for new algorithmic
ideas and data reduction procedures~\cite{FerizovicHLM0S20,FigielFNN22,HespeL0S20,Hespe0S19}. 

Many problems that have been studied thoroughly in the literature
turn out to be vertex deletion problems. The simplest example arises from
\emph{vertex covers,} which measure the ``distance in terms of vertex
deletions'' of a graph from being edge-free:  A graph has a vertex
cover of size~$k$ iff it can be made edge-free by deleting at most
$k$~vertices. For a slightly more complex example, the \emph{cluster
deletion problem} asks whether we can delete at most~$k$ 
vertices from a graph so that it becomes a cluster graph, meaning
that every connected component is a clique or, equivalently, is
$P_3$-free (meaning, there is no induced path on three
vertices). The \emph{feedback vertex set problem} asks if we can
delete at most~$k$ vertices, such that the resulting graph has no
cycles.  The \emph{odd cycle transversal problem} asks if there is a
set of vertices of size at most $k$, such that removing it destroys
every odd  cycle. Equivalently, the problem asks if we can delete at
most $k$ vertices, such that the resulting graph is bipartite.

To investigate
the complexity of vertex deletion problems in a systematic way, it makes sense
to limit the graph properties to have some structure. An early result
in this direction~\cite{LewisY80} is the $\Class{NP}$-completeness of vertex deletion
to hereditary graph properties that can be tested in polynomial
time. Intuitively, vertex deletion problems should be easier to
solve for graph properties that are simpler to express. Phrased in
terms of descriptive complexity theory, if we can describe a graph
property using, say, a simple first-order formula, the corresponding
vertex deletion problem should also be simple. The intuition was
proven to be correct in 2020, when Fomin et~al.~\cite{FominGT20}
established a dichotomy based on the number  of quantifier
alternations that characterizes the classes of first-order logic
formulas for which the vertex deletion problem is fixed-parameter
tractable.

The results of Fomin et al.\ directly apply to some of the above
examples: Consider the problem $\PLang{vertex-cover}$, whose
``triviality'' property is described by the formula $\phi_\text{vc} = \forall
x\forall y (x\nadj y)$, or the problem $\PLang{cluster-deletion}$,
whose triviality property is described by $\phi_\text{cd} = \forall
x\forall y\forall z\bigl( (x\adj y \land
y\adj z) \to x\adj z\bigr)$. Both first-order formulas use \emph{no
quantifier alternations,} which by~\cite{FominGT20} already implies
that the problems lie in $\Para\Class{P} = \Class{FPT}$. Naturally,
not all problems can be characterized so easily: Properties like
acyclicity (which underlies the feedback vertex set problem) cannot be
expressed in first-order logic and, thus, the results of Fomin
et~al.\ do not apply to them. Fomin et~al.\ also show that if there
are enough quantifier alternations (three, to be precise) in the
first-order formulas describing the property, then the resulting
vertex deletion problem can be $\Class W[1]$-hard. Nevertheless, the
descriptive approach allows us to identify large fragments of logical
formulas and hence large classes of vertex deletion problems that are
(at least fixed-parameter) tractable. 

A first central question addressed in the present paper is whether the
number of quantifier \emph{alternations} (the property studied
in~\cite{FominGT20}) overshadows all other aspects in  making problems
hard, or whether the individual quantifier pattern of the formula
plays a significant role as well. This question appears to be of
particular importance given that formulas describing natural problems
(like $\phi_\text{vc}$ and $\phi_\text{cd}$ above) tend to have short
and simple quantifier patterns: We might hope that even though we
describe a particular triviality property using, say, four alternations,
the fact that we use only, say, two existential quantifiers in total
still assures us that the resulting vertex deletion problem is easy.

A second central question is whether the \emph{kind} of graphs that we
allow as inputs has an influence on the complexity of the
problem. Intuitively, allowing only, say, \emph{basic graphs}
(simple undirected graphs without self-loops) should result in simpler
problems than allowing directed graphs or even arbitrary logical
structures as input. This intuition is known to be correct in the
closely related question of deciding graph properties described in existential second-order logic. As we will see, in the
context of vertex deletion problems it makes a difference whether we
consider basic graphs, undirected graphs, or directed graphs, but not
whether we consider directed graphs or arbitrary logical structures.

\subparagraph*{Our Contributions.}

We completely classify the parameterized complexity of vertex
deletion problems in dependence of the quantifier pattern of the
formulas that are used to express the triviality property and also in
dependence of the kind of graphs that we allow as inputs (basic,
undirected, directed, or arbitrary logical structures). An overview of
the results is given in Table~\ref{table:summary}, where the following
notations are used (detailed definitions are given later): For a
first-order formula~$\phi$ over the vocabulary $\tau = \{\sim^2\}$ of
(directed, simple) graphs, the parameterized problem 
$\PLang[k]{vertex-deletion}_{\mathrm{dir}}(\phi)$ (abbreviated
$\PLang{vd}_{\mathrm{dir}}(\phi)$) asks us to tell on input of a
directed graph~$G$ and a parameter $k\in\mathbb{N}$ whether
we can delete at most $k$ vertices from~$G$, so that for the resulting
graph~$G'$ we have $G'\models\phi$.
The problems
$\PLang{vd}_{\mathrm{undir}}(\phi)$ and
$\PLang{vd}_{\mathrm{basic}}(\phi)$ are the restrictions where the
input graphs are undirected or basic graphs
(undirected graphs without self-loops), respectively.  For instance,
$\PLang{vertex-cover} = \PLang{vd}_{\mathrm{basic}}(\phi_{\text{vc}}) =
\PLang{vd}_{\mathrm{basic}}\bigl(\forall x\forall y (x\nadj y)\bigr)$. 
In the other direction, let $\PLang{vd}_{\mathrm{arb}}(\phi)$ denote the
generalization where we allow an arbitrary logical vocabulary~$\tau$ and
arbitrary (finite) logical structures $\mathcal A$ instead of just
graphs~$G$ (and where ``vertex deletion'' should better be called
``element deletion,'' but we stick with the established name).
For a \emph{(first-order) quantifier pattern~$p$,} which is just a
string of $a$'s and $e$'s standing for the universal and existential
quantifiers at the beginning of a formula~$\phi$, we write
$\PVD{basic}(p)$ for the class of all problems
$\PLang{vd}_{\mathrm{basic}}(\phi)$ where $\phi$ has all its quantifiers at the 
beginning and they form the pattern~$p$. For
instance, $\PLang{vertex-cover} \in \PVD{basic}(aa)$ as $\phi_{\text{vc}}$
has two universal quantifiers. The same notation is used for
undirected graphs, directed graphs, and arbitrary structures. 

\begin{table}[htbp]
  \caption{Complete complexity classification of vertex deletion
    problems for first-order formulas in dependence of the quantifier
    pattern $p \in \{a,e\}^*$ (where $p \preceq q$ means that $p$ is a 
    subsequence of~$q$). The four different considered restrictions on the
    allowed input structures lead to three distinct complexity
    landscapes. Note that $\Para\Class{AC}^0 
    \subsetneq \Para\Class{AC}^{0\uparrow} \subseteq \Para\Class P = \Class{FPT}$
    holds and that it is a standard assumption that $\Class{FPT} \cap
    \Class W[2]\text{-hard} =\emptyset$ also holds.}
  \label{table:summary}
  \begin{tabular}{llrcl}
    \toprule
    $\PVD{basic}(p)$
    & $\subseteq \Para\Class{AC^{0}}$, when 
    &
    & $\AtMost$
    & $e^*a^*$ or $eae$.
    \\
    & $\not\subseteq \Para\Class{AC^{0}}$ but $\subseteq \Para\Class{AC^{0\uparrow}}$, when
    & $eeae$, $aae$ or $aee$
    & $\Between$
    & $e^*a^*e^*$.
    \\
    &  $\cap\ \Class{W}[2]\text{-hard} \neq \emptyset$, when
    & $aea$
    & $\AtLeast$
    &
    \\
    \midrule
    $\PVD{undir}(p)$
    & $\subseteq \Para\Class{AC^{0}}$, when
    &
    & $\AtMost$
    & $ae$ or $e^*a^*$.
    \\
    & $\not\subseteq \Para\Class{AC^{0}}$ but $\subseteq \Para\Class{AC^{0\uparrow}}$, when
    & $eae$, $aae$ or $aee$
    & $\Between$
    & $e^*a^*e^*$.
    \\
    & $\cap\ \Class{W}[2]\text{-hard} \neq \emptyset$, when
    & $aea$
    & $\AtLeast$
    &
    \\
    \midrule
    $\PVD{dir}(p)$ and
    & $\subseteq \Para\Class{AC^{0}}$, when
    &
    & $\AtMost$
    & $e^*a^*$.
    \\
    $\PVD{arb}(p)$ & $\not\subseteq \Para\Class{AC^{0}}$ but $\subseteq \Para\Class{AC^{0\uparrow}}$, when
    & $ae$
    & $\Between$
    & $e^*a^*e^*$.
    \\
    &  $\cap \ \Class{W}[2]\text{-hard} \neq \emptyset$, when
    & $aea$
    & $\AtLeast$
    &
    \\
    \bottomrule
  \end{tabular}
\end{table}

The results in Table~\ref{table:summary} give an answer to the first 
central question formulated earlier, which asked whether it is the
\emph{number of alternations} of quantifiers in patterns (and not so
much the actual number of quantifiers) that are responsible for the
switch from tractable to intractable observed by Fomin
et~al.~\cite{FominGT20}, or whether the frontier is formed by short
patterns that ``just happen'' 
to have a certain number of alternations. As can be seen, the latter
is true: All intractability results hold already for very short and
simple patterns. Thus, while it was previously known that
there is a formula in $\Pi_3$ (meaning it has a pattern of the form
$\forall^*\exists^*\forall^*$ or $a^*e^*a^*$ in our notation) defining
an intractable problem, we show that already one quantifier per
alternation (the pattern $aea$) suffices. 
%
%
On the positive side, Table~\ref{table:summary} shows
that all vertex deletion problems that are (fixed-parameter) tractable
at all already lie in the classes $\Para\Class{AC^{0}}$ or at least
$\Para\Class{AC}^{0\uparrow}$. From an algorithmic point of view this
means that all of the vertex deletion problems that we classify as 
fixed-parameter tractable admit efficient \emph{parallel}
fixed-parameter algorithms.

Concerning the second central question, which asked whether it makes a
difference which kind of graphs or logical structures we consider,
Table~\ref{table:summary} also provides a comprehensive answer: First,
the \emph{frontier of tractability} (the patterns where we switch from
membership in $\Class{FPT} = \Para\Class P$ to hardness for
$\Class{W}[1]$) \emph{is the same for all kinds of inputs} (namely
from ``does not contain $aea$ as a subsequence'' to ``contains $aea$
as a subsequence''). Second, if we classify the tractable fragments
further according to ``how tractable'' they are, a
more complex complexity landscape arises: While $\PVD{dir}(p)$ and
$\PVD{arb}(p)$ have the same classification for all~$p$, the classes
$\PVD{basic}(p)$ and $\PVD{undir}(p)$ each exhibit a different
behavior. In other words: For simple patterns~$p$, it makes a
difference whether the inputs are basic, undirected, or directed graphs.

The just-discussed structural results are different from
classifications in dependence of quantifier patterns~$p$ 
established in previous works: Starting with 
Eiter et~al.~\cite{EiterGG00} and subsequently Gottlob
et~al.~\cite{GottlobKS04}, Tantau~\cite{Tantau15} and most recently 
Bannach et al.~\cite{BannachCT23}, different authors have classified
the complexity of \emph{weighted definability problems} by the 
quantifier patterns used to describe them. In these problems, formulas
have a free set variable and we ask whether there is an assignment to
the set variable with at most $k$~elements such that the formula  
is true. Since it is easy to see that the vertex deletion problems we
study are special cases of this question, upper
bounds from earlier research also apply in our setting. However, our
results show that (as one would hope) for vertex deletion problems for
many patterns~$p$ we get better upper bounds than in the more general
setting. Furthermore, there is an interesting structural insight
related to our second central question: While the results
in~\cite{BannachCT23} for weighted definability show that, there, the
complexities for undirected graphs, directed graphs, and
arbitrary logical structures all coincide (but differ for basic
graphs), for the vertex deletion setting, we get three different
complexity characterizations for basic, undirected, and directed
graphs -- but the latter coincide with arbitrary structures once more.

\subparagraph*{Related Work.}

The complexity-theoretic investigation of vertex deletion
problems has a long and fruitful history. Starting in classical complexity
theory, results on vertex deletion problems were established as early as in the late
1970s~\cite{Krishnamoorthy79, LewisY80, Yannakakis78}. The focus was mostly on
deletion to commonly known graph properties, such as planarity, acyclicity or
bipartiteness.

Since it is very natural to regard the number of allowed modifications as the parameter of the
problem, the investigation of vertex deletion problems quickly gained traction
in parameterized complexity, with continued research to this
day~\cite{ChenLLOR08, FominLPSZ20, LiN20}. Specifically for graphs, similar problems like
the deletion or modification of edges~\cite{CRESPELLE2023100556} or alternative
distance measures such as elimination distance~\cite{FominGT21} are also
considered. Regarding first-order definable properties, a dichotomy is shown
in~\cite{FominGT20}.

The framework of quantifier prefix patterns we employ in this paper has also
received a lot of attention, especially in the context of descriptive complexity.
Early uses go as far back as the classification of decidable fragments of
first-order logic~\cite{BorgerGG1997}. They were then considered in the context of classical
complexity~\cite{EiterGG00,GottlobKS04,Tantau15} and later also in the context
of parameterized complexity~\cite{BannachCT23}.

\subparagraph*{Organization of this Paper.}

Following a review of basic concepts and terminology in
Section~\ref{section:background}, we present the complexity-theoretic
classification of the vertex deletion problems for basic, undirected and
directed graphs in Sections~\ref{section:basic-graphs},
\ref{section:undirected-graphs} and \ref{section:directed-graphs}, respectively.
\tcsautomoveaddto{main}{
  \clearpage
  \appendix
  \section{Technical Appendix}
  In the following, we provide the proofs omitted in the main text. In
  each case, the claim of the theorem or lemma is stated once more for
  the reader's convenience. 
}

\section{Background in Descriptive and Parameterized Complexity}
\label{section:background}
 

\subparagraph*{Terminology from Finite Model Theory.}

In this paper, we will use standard terminology from finite model
theory, for a thorough introduction, see, for
example~\cite{EbbinghausF05}. A \emph{relational vocabulary}~$\tau$
(also known as a \emph{signature}) is a set of \emph{relation symbols}
to each of which we assign a positive \emph{arity,} denoted using a
superscript. For example, $\tau = \{P^1, E^2\}$ is a relational
vocabulary with a monadic relation symbol $P$ and a dyadic relation
symbol~$E$. A \emph{$\tau$-structure~$\mathcal A$} consists of a
\emph{universe~$A$} and for each relation symbol $R \in \tau$ of some
arity~$r$ of a relation $R^{\mathcal A} \subseteq A^r$.  We denote the
set of \emph{finite $\tau$-structures} as $\Lang{struc}[\tau]$. For
a first-order $\tau$-sentence $\phi$, we write $\Lang{models}(\phi)$
for the class of finite models of $\phi$.
A \emph{decision problem}~$P$ is a subset of $\Lang{struc}[\tau]$
which is closed under isomorphisms. A formula $\phi$ \emph{describes}
$P$ if $\Lang{models}(\phi) = P$. 

For $\tau$-structures
$\mathcal{A}$ and $\mathcal B$ with universes $A$ and~$B$,
respectively, we say that $\mathcal{A}$ is an \emph{induced
substructure} of $\mathcal{B}$ if $A \subseteq B$ and for all $r$-ary 
$R\in \tau$, we have $R^{\mathcal{A}} = R^{\mathcal{B}} \cap A^r$.
For a set $S\subseteq B$, we denote by $\mathcal{B}\setminus S$ the
substructure induced on $B \setminus S$. 

We regard \emph{directed} graphs $G = (V, E)$ (which are pairs of a
nonempty vertex set~$V$ and an edge relation $E \subseteq V \times V$)
as logical structures $\mathcal G$ over the vocabulary
$\tau_{\text{digraph}} = \{\adj^2\}$ where $V$ is the universe and
$\adj^{\mathcal G} = E$. An \emph{undirected} graph is a directed
graph that additionally satisfies $\phi_{\text{undirected}} \coloneq
\forall x\forall y (x\adj y \to y\adj x)$, while a \emph{basic} graph
satisfies $\phi_{\text{basic}} \coloneq \forall x\forall y \bigl(x\adj
y \to (y\adj x \land x \neq y)\bigr)$.  

For a first-order logic formula in prenex normal form (meaning all
quantifiers are at the front), we can associate a \emph{quantifier prefix  
pattern} (or \emph{pattern} for short), which are words over the alphabet
$\{e, a\}$.\footnote{One uses ``$a$'' and ``$e$'' in patterns rather than
``$\forall$'' and ``$\exists$'' since in the context of second-order
logic one needs a way to differentiate between first-order
and second-order quantifiers and, there, ``$E$'' refers to a
``second-order $\exists$'' while ``$e$'' refers to a ``first-order
$\exists$''. In our paper, we only use first-order quantifiers so only
lowercase letters are needed.}
For example, the formula $\phi_{\text{basic}}$ has the
pattern~$aa$, while the formula $\phi_{\text{degree-$\geq$2}} \coloneq
\forall x \exists y_1 \exists y_2 \bigl((x \adj y_1) \land (x\adj y_2)
\land (y_1 \neq y_2)\bigr)$ has the pattern~$aee$. As another example,
the formulas in the class~$\Pi_2$ (which start with a universal
quantifier and have one alternation) are exactly the formulas with a
pattern $p \in \{a\}^* \circ \{e\}^*$, which we write briefly as $p
\in a^*e^*$. We write $p \preceq q$ if $p$ is a subsequence of $q$.

\subparagraph*{Terminology from Parameterized Complexity.}

We use standard definitions from parameterized complexity, see for
instance~\cite{Cygan15, DowneyF99, FlumG06}. A \emph{parameterized
problem} is a set $Q\subseteq \Sigma^* \times \mathbb{N}$ for an
alphabet~$\Sigma$. In an \emph{instance} $(x,k) \in \Sigma^* \times \mathbb{N}$
we call $x$ the \emph{input} and $k$ the \emph{parameter}. The central
problem we consider in this paper is the following: 

\begin{problem}[{$\PLangVD{arb}(\phi)$, where $\phi$ is a first-order $\tau$-formula}]
  \begin{parameterizedproblem}
    \instance (An encoding of) a logical $\tau$-structure $\mathcal{A}$ and an integer $k\in\mathbb{N}$.
    \parameter $k$.
    \question Is there a set $S\subseteq A$ with $|S| \leq k$ such that
    $\mathcal{A}\setminus S\models\phi$?
  \end{parameterizedproblem}
\end{problem}
%
As mentioned earlier, we also consider the problems
$\PLangVD{basic}(\phi)$, where the input structures are basic graphs
(formally, $\PLangVD{basic}(\phi) = \PLangVD{arb}(\phi) \cap
\bigl(\Lang{models}(\phi_{\text{basic}})\times\mathbb N\bigr)$), the
problems $\PLangVD{undir}(\phi)$, where the input structures are undirected
graphs, and $\PLangVD{dir}(\phi)$, where the input 
structures are directed graphs.
For a pattern $p \in \{a,e\}^*$, the class $\PVD{arb}(p)$
contains all problems $\PLangVD{arb}(\phi)$ such that $\phi$ has
pattern~$p$. The classes with the subscripts ``basic'', ``undir'', and
``dir'' are defined similarly.



We will consider some parameterized circuit complexity classes. We define $\Para\Class{AC}^0$ as the class of
parameterized problems that can be decided by a family of unbounded fan-in circuits $(C_{n,
k})_{n, k\in \mathbb{N}}$ of constant depth and size $f(k) \cdot n^{O(1)}$ for
some computable function $f$. Similarly, $\Para\Class{FAC}^0$ is the class of
functions that can be computed by a family of unbounded fan-in circuits $(C_{n,
k})_{n, k\in \mathbb{N}}$ of constant depth and size $f(k) \cdot n^{O(1)}$ for
some computable function $f$. For
$\Para\Class{AC}^{0\uparrow}$, we allow the circuit to have depth $f(k)$.
Questions of uniformity will not be important in the present paper. For
these classes, we have the following inclusions: $\Para\Class{AC}^0 \subsetneq
\Para\Class{AC}^{0\uparrow} \subseteq \Para\Class{P} = \Class{FPT}$.

A parameterized problem $Q\subseteq \Sigma^* \times \mathbb{N}$ is
$\Para\Class{AC}^{0}$-many-one-reducible to a problem $Q'\subseteq
\Gamma^* \times \mathbb{N}$, written $Q
\le^{\Para\Class{AC}^0}_{\mathrm{m}} Q'$, if there is a function
$f\colon \Sigma^* \times \mathbb{N} \to \Gamma^* \times \mathbb{N}$, such that
(1) for all $(x, k) \in \Sigma^* \times \mathbb{N}$ we have $(x, k) \in Q$ iff
$f(x, k) \in Q'$, (2) there is a computable function $g\colon
\mathbb{N} \to \mathbb{N}$ such that for all $(x, k) \in \Sigma^*
\times \mathbb{N}$, we have $k' \leq g(k)$, where $f(x, k) = (x',
k')$, and (3) $f\in \Para\Class{FAC}^0$. The more general
$\Para\Class{AC}^{0}$ disjunctive truth table reduction, 
written $Q \le^{\Para\Class{AC}^0}_{\mathrm{dtt}} Q'$, is defined
similarly, only $f$ maps $(x,k)$ to a sequence
$(x_1,k_1),\dots,(x_\ell,k_\ell)$ of instances such that (1$'$) $(x,k) \in Q$ iff there is an $i \in \{1,\dots,\ell\}$ with $(x_i,k_i) \in Q'$ and (2$'$) $k_i \le g(k)$ holds for all $i \in
\{1,\dots,\ell\}$.   
Both $\Para\Class{AC}^{0}$ and $\Para\Class{AC}^{0\uparrow}$ are closed under
$\le^{\Para\Class{AC}^0}_{\mathrm{m}}$- and $\le^{\Para\Class{AC}^0}_{\mathrm{dtt}}$-reductions.

%
%
\section{Basic Graphs}
\label{section:basic-graphs}

\tcsautomoveaddto{main}{\subsection{Proofs for Section~\ref{section:basic-graphs}}}

Basic graphs, that is, undirected graphs without self-loops, are one of the
simplest non-trivial logical structures one can imagine. Despite that, many
$\Class{NP}$-hard problems on graphs, like vertex cover, clique or
dominating set, are $\Class{NP}$-hard even for basic graphs.
%
This also transfers in some sense to our setting: The ``tractability frontier'',
the dividing line between the fragments which are tractable and those
where we can express intractable problems, is the same for all graph classes we consider.
%
However, when we shift our attention to the complexity landscape inside the tractable
fragments, we also see that the complexity of the logical structure has an
impact on the complexity of the problems we can define: Basic, undirected, and
directed graphs all have provably distinct complexity characterizations.

We begin by stating the main theorem of the section, the complexity
classification for basic graphs. In the rest of the section, we show the upper
and lower bounds that lead to this classification.

\begin{theorem}[{Complexity Trichotomy for $\PVD{basic}(p)$}]\label{theorem:basic}
  Let $p \in \{a,e\}^*$ be a pattern.
  \begin{enumerate}
  \item $\PVD{basic}(p) \subseteq \Para\Class{AC}^{0}$, if $p
    \preceq eae$ or $p \preceq e^*a^*$.
  \item $\PVD{basic}(p) \subseteq \Para\Class{AC}^{0\uparrow}$
    but $\PVD{basic}(p) \not\subseteq \Para\Class{AC}^{0}$, if
    $eeae \preceq p$, $aae \preceq p$ or $aee \preceq p$ holds, but also still $p
    \preceq e^*a^*e^*$.
  \item $\PVD{basic}(p)$ contains a $\Class{W[2]}$-hard
    problem, if $aea\preceq p$.
  \end{enumerate}
\end{theorem}

The theorem covers all possible patterns. It follows from the following lemma,
where we state the individual complexity characterizations we will prove:

\begin{lemma}[Detailed Bounds for $\PVD{basic}(p)$]\label{lemma:trichotomy-basic}\hfil
  \begin{enumerate}
    \item $\PVD{basic}(eae) \subseteq \Para\Class{AC}^{0}$.
    \item $\PVD{basic}(e^*a^*) \subseteq\PVD{arb}(e^*a^*) \subseteq \Para\Class{AC}^{0}$.
    \item $\PVD{basic}(e^*a^*e^*) \subseteq \PVD{arb}(e^*a^*e^*) \subseteq\Para\Class{AC}^{0\uparrow}$.
    \item $\PVD{basic}(eeae)$ contains a problem not in $\Para\Class{AC}^{0}$.
    \item $\PVD{basic}(aae)$ contains a problem not in $\Para\Class{AC}^{0}$.
    \item $\PVD{basic}(aee)$ contains a problem not in $\Para\Class{AC}^{0}$.
    \item $\PVD{basic}(aea)$ contains a $\Class{W}[2]$-hard problem.
  \end{enumerate}
\end{lemma}

Notice that in particular, we know unconditionally that
$\Class{W}[2]\not\subseteq \Para\Class{AC}^{0}$, and, hence, a
$\Class{W}[2]$-hard problem cannot lie in $\Para\Class{AC}^{0}$. It is furthermore widely conjectured
that $\Class{W}[2]\not\subseteq \Para\Class{AC}^{0\uparrow}$, as
$\Para\Class{AC}^{0\uparrow}\subseteq \Class{FPT}$. We devote the rest of
this section to proving the individual items of the lemma.


\subparagraph*{Upper Bounds}
Previous work by Bannach et al.~\cite{BannachCT23}
showed that in the weighted definability setting, formulas
with the pattern~$ae$ already suffice to describe $\Class{W}[2]$-hard
problems. We now show that the situation is more favorable in the
vertex deletion setting, which is a special case of weighted
definability: All problems in $\PVD{basic}(e^*a^*e^*)$ are tractable
and the problems in $\PVD{basic}(e^*a^*)$ and in $\PVD{basic}(eae)$ 
are even in $\Para\Class{AC}^0$, the smallest class commonly
considered in parameterized complexity. We start with the last claim:

\begin{lemma~}\label{lemma:vd-basic-eae}
  $\PVD{basic}(eae) \subseteq \Para\Class{AC}^{0}$.
\end{lemma~}

\begin{proof}[Proof idea]
  To check
  whether we can delete at most $k$~vertices to satisfy a formula with prefix
  pattern $eae$, we first branch over the possible assignments to the first
  existentially quantified variable. Now, the neighborhood of this variable
  induces a $2$-coloring on the rest of the graph. For the rest of the prefix,
  $ae$, we prove that a vertex has to be deleted if and only if there is no
  special set of constant size, called \emph{stable set}. This can all be checked in $\Para\Class{AC}^{0}$.
\end{proof}

\begin{proof~}
  Fix a formula $\phi$ with pattern~$eae$. Then we can rewrite $\phi$
  equivalently in the following form for some quantifier-free formulas
  $\phi_1$ and~$\phi_2$, neither of which contains the atoms $s = x$
  or $s\neq x$:
  \begin{align}
    \exists s\forall x\exists y\bigl(((s = x) \to \phi_1(s, x,
    y)) \land ((s \neq x) \to \phi_2(s, x, y))\bigr). \label{eq-phi'}
  \end{align}


  We wish to show that $\PVD{basic}(\phi)$ can be decided by a
  $\Para\Class{AC}^{0}$ algorithm. Let $G = (V,E)$ be an input graph
  for our algorithm. 

  We start with some terminology: Since the formula asks us to find
  for all vertices $x \in V \setminus \{s\}$ a vertex $y \in V$ such
  that   
  $\phi_2(s,x,y)$ holds, we call such a $y$ a \emph{witness
  for~$x$ (relative to~$s$)}. We denote by~$W_x^s$  the set of possible
  witnesses for~$x$ relative to~$s$ and note that $s \in W_x^s$ may
  hold. Observe that when the existential quantifier 
  $\exists s$ is instantiated with some particular value $s \in V$ and
  if $W_x^s = \emptyset$ holds, we \emph{have to} delete~$x$ to make
  the rest of the formula (the part following $\exists s$) true. A
  \emph{witness walk (relative to~$s$) starting at~$v_1$} or just a
  \emph{$v_1$-witness walk} is a sequence of vertices $(v_1, v_2,
  \dots, v_j)$ such that 
  \begin{enumerate}
  \item we have $v_{i + 1}\in   W_{v_{i}}^s$ for all $i\in\{1, \dots,j-1\}$,
  \item the vertices $v_1$ to $v_{j-1}$ are distinct, and
  \item we have $v_j = s$ (and say that the the walk is \emph{$s$-terminated}) or
    $v_j =  v_i$ for some $i < j$ (and say that the walk is \emph{returning}) or
    $W_{v_j}^s = \emptyset$ (and say that the walk is \emph{unstable}).
  \end{enumerate}
  A walk is \emph{stable} if it is not unstable (so it is $s$-terminated or
  returning). Our first crucial observation is that we never have to
  delete vertices that are part of a stable walk to make the graph
  satisfy the formula. Formally:


    

  \begin{claim}\label{claim:stable-or-delete}
    Fix $s \in V$. Then for every vertex $v \in V \setminus \{s\}$
    there is either a stable $v$-witness walk (relative to~$s$) or $v$
    has to be deleted in order to satisfy~$\phi$ when the
    existential quantifier is instantiated with the fixed~$s$.  
  \end{claim}

  \begin{proof}
    Suppose that there is no stable $v$-witness walk. Consider the
    $v$-witness walk obtained by arbitrarily adding consecutive
    witnesses to the walk as long as possible. As this walk is
    unstable, it ends with a vertex $v_j \neq s$ with $W_{v_j}^s =
    \emptyset$. Thus, there is no way to make $\exists y
    \phi_2(s,v_j,y)$ true in~$G$ and thus also not $\forall x \exists
    y \phi_2(s,x,y)$. In particular, we need to delete~$v_j$, making
    the graph smaller, and note that this does not introduce any stable
    $v$-witness walks. Thus, by repeating the argument often enough,
    at some point we must have $j= 1$, that is, $v_1 = v_j$ and
    $W_{v_1}^s = \emptyset$ holds. This means that we must delete $v$
    in order to make $\phi$ true, as claimed.
  \end{proof}


  By the claim, for each fixed $s\in V$, we \emph{have to delete
  the vertices that are not the starts of stable witness walks
  to make the graph satisfy the formula} and also note that \emph{we
  do not have to delete vertices $v$ for which a stable $v$-witness
  walk exists as each vertex on it has a witness}. Since we will soon
  see that it suffices to consider stable witness walks of
  length~$10$, we get the following algorithm: 

\input{basic-eae-algo.tex}

  The algorithm can be implemented in $\Para\Class{AC}^{0}$: For the
  for-statement in line~\ref{line-branch} we branch
  over the possible choices of~$s$ using $|V|$ copies of the circuit
  executing the rest of the algorithm. Finding a stable $v$-witness
  walk for each vertex~$v$ in line~\ref{line-stable} can be done 
  using $|V|^{10}$ parallel subcircuits. We can implement the handling of
  the set~$D$ by encoding it using a bit vector of length $|V|$ where
  the $i$th bit is set when the $i$th vertex is in~$D$: This allows us
  to add vertices to~$D$ in line~\ref{line-add} in constant depth, and
  it is known~\cite{BannachST15,BannachT18,ChenF16} that the size
  check $|D| \le k$ in line~\ref{line-size-check} can be implemented
  in $\Para\Class{AC}^{0}$. The final check ``$G' \models \exists
  y(\phi_1(s, s, y))$'' in line~\ref{line-final-check} can trivially
  be done using an $\Class{AC}^0$ circuit as $\phi_1$ is a first-order
  formula.  
  
  We show the correctness of the algorithm in two directions: For the first
  direction, observe that if the algorithm outputs that $(G,k)$ lies in
  $\PLangVD{basic}(\phi)$ in line~\ref{line-accept}, we have just
  found a vertex $s\in V$ and a set~$D$ of at most $k$ vertices whose
  deletion yields a graph $G'$ that satisfies the formula. This is
  because $G'$ satisfies $\forall x \exists y ((s = x) \to \phi_1(s,
  x, y))$ (since this is equivalent to $\exists y \phi_1(s,s,y)$ and
  we have just tested this in line~\ref{line-final-check}) and every
  vertex $v \in V \setminus \{s\}$ has a witness (since there is a
  stable $v$-witness walk we know that each vertex on it has a
  witness and no vertex on it ever becomes part of~$D$), so $G'$ also
  satisfies $\forall x \exists y((s \neq x) \to \phi_2(s, x, y))$. All
  told, $G'$ is a model of~\eqref{eq-phi'}.  
  
  For the other direction, we show that if we have $(G,k) \in
  \PLangVD{basic}(\phi)$, then the algorithm outputs this in
  line~\ref{line-accept}. Membership in $\PLangVD{basic}(\phi)$
  implies that there is a $s \in V$ and at least one set
  $D^s \subseteq V \setminus\{s\}$ with $|D^s| \le k$ such that 
  \begin{align} 
    G \setminus D^s \models \forall x\exists y\bigl(((s = x) \to \phi_1(s,
    x, y)) \land ((s \neq x) \to \phi_2(s, x, y))\bigr). \label{eq-d}
  \end{align}
  The algorithm will consider this particular $s$ at some point in
  line~\ref{line-branch}. We show in a moment that in lines~\ref{line-d-start}
  to~\ref{line-d-end} the algorithm then computes exactly the smallest
  set~$D$ that makes \eqref{eq-d} hold. In particular, this implies
  that $|D| \le k$ will hold, which in turn means that the test in
  line~\ref{line-size-check} is passed and so is the final check in
  line~\ref{line-final-check} as \eqref{eq-d} holds. Thus, the
  algorithm will produce the correct output in line~\ref{line-accept}
  as claimed.

  To show that the minimal~$D$ is computed, first note that by
  Claim~\ref{claim:stable-or-delete} we \emph{have to} delete all
  vertices that are not part of stable sets. Thus, if we can prove
  that we put exactly the vertices $v \in V \setminus \{s\}$ into~$D$
  for which there is no stable $v$-witness walk, we are done: We have to delete
  all of them, but we delete no more and as part of stable witness
  walks, all remaining vertices~$x$ have a witness. However, in
  line~\ref{line-stable} we only check whether there is a stable
  $v$-witness walk \emph{of length~$10$} and it remains to prove
  that this test is sufficient. That is, we have to show that for
  every vertex $v \in V \setminus \{s\}$ for which there is some
  stable $v$-witness walk, there is also one of length at
  most~$10$. This is exactly the final claim:
  
  \begin{claim}\label{claim:stable-set-size}
    For each $v \in V \setminus \{s\}$, if there is a stable
    $v$-witness walk relative to~$s$, there is also one of length at
    most~$10$. 
  \end{claim}

  \begin{proof}
    Consider a shortest stable $v$-witness walk
    $(v_1,v_2,v_3,\dots,v_j)$ relative to~$s$ that starts at
    $v_1=v$.
    We wish to show $j \le 10$, so for the sake of 
    contradiction assume $j > 10$. Then none of $v_1$ to $v_{10}$ can
    equal~$s$ and all of them must be distinct.
    
    Recall that a witness of a vertex $x \in V \setminus \{s\}$ is a
    vertex $y \in V$ such that $\phi_2(s,x,y)$ holds. We may assume
    that the formula $\phi_2$ contains as its atomic formulas only
    $x=y$, $s=y$, $x\adj y$, $x\adj s$, $y\adj s$, as well as
    negations thereof, since $\phi_2$ is guarded by ``$s\neq x\to$''
    inside \eqref{eq-phi'} and since $x \adj x$, $y \adj y$, and
    $s\adj s$ are all always false in basic graphs. Furthermore,
    whether or not $\phi_2(s,v_p,v_q)$ holds for $p,q \in
    \{1,\dots,10\}$ with $p\neq q$ depends neither on the
    atoms $x = s$ nor on $x = y$ inside~$\phi_2$, since, should these be
    present, they will always be false. Rather, the only remaining
    atoms that can still be relevant inside $\phi_2$ are $x\adj y$,
    $x\adj s$, and $y\adj s$. 

    Let us say that a vertex is \emph{black} if it is adjacent to~$s$,
    otherwise it is \emph{white.} Then, for all $p,q \in \{1,\dots,10\}$
    with $p \neq q$, the question of whether $v_q$ is a witness
    for~$v_p$ relative to~$s$ (that is, whether  $\phi_2(s,v_p,v_q)$
    holds), depends only on whether $v_q \adj v_p$ holds and on the
    colors of $v_p$ and~$v_q$.

    We now distinguish two cases: First, that the stable $v_1$-witness
    walk $(v_1,v_2,v_3,\dots,v_j)$ is $s$-terminated (meaning $v_j =
    s$) or is returning to $v_i$ with $i \ge 5$. Second, that the
    witness walk is returning, but to some $v_i$ with $i < 5$.

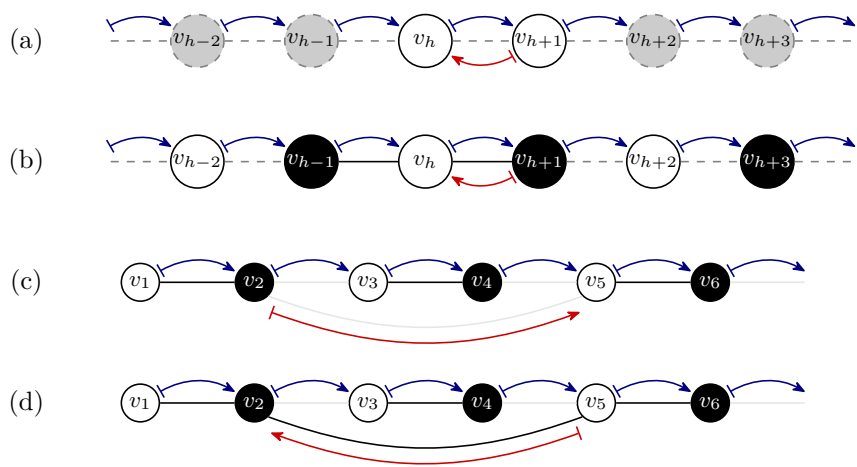
\begin{figure}[htpb]
  \centering
  \begin{tikzpicture}[
      x=1.5cm,
      y=0.8cm
    ]
    
    \begin{scope}[
        node/.append style={minimum size=7mm, inner sep=0pt}
      ]
      \begin{scope}[shift={(0.5,0)}]
        \node (hm3) at (-1,0) [virtual node] {};
        \node (hm2) at (0,0) [gray node]  {$v_{h-2}$};
        \node (hm1) at (1,0) [gray node]  {$v_{h-1}$};
        \node (h)   at (2,0) [white node] {$v_h$};
        \node (hp1) at (3,0) [white node] {$v_{h+1}$};
        \node (hp2) at (4,0) [gray node]  {$v_{h+2}$};
        \node (hp3) at (5,0) [gray node]  {$v_{h+3}$};
        \node (hp4) at (6,0) [virtual node] {};
        
        \foreach \from/\to in {hm3/hm2,hm2/hm1,hm1/h,h/hp1,hp1/hp2,hp2/hp3,hp3/hp4} {
          \draw [arbitrary] (\from) -- (\to);
          \draw [mapsto] (\from) to [bend left=30] (\to);
        }
        
        \draw [mapsto,hilight] (hp1) to [bend left=30] (h);
      \end{scope}
          
      \node at (-1,0) {(a)};
    \end{scope}

    \begin{scope}[
        shift={(0,-2)},
        node/.append style={minimum size=7mm, inner sep=0pt}
      ]
      \begin{scope}[shift={(0.5,0)}]
        \node (hm3) at (-1,0) [virtual node] {};
        \node (hm2) at (0,0) [white node]  {$v_{h-2}$};
        \node (hm1) at (1,0) [black node]  {$v_{h-1}$};
        \node (h)   at (2,0) [white node] {$v_h$};
        \node (hp1) at (3,0) [black node] {$v_{h+1}$};
        \node (hp2) at (4,0) [white node]  {$v_{h+2}$};
        \node (hp3) at (5,0) [black node]  {$v_{h+3}$};
        \node (hp4) at (6,0) [virtual node] {};
        
        \foreach \from/\to/\how in {hm3/hm2/arbitrary,hm2/hm1/arbitrary,hm1/h/present,h/hp1/present,hp1/hp2/arbitrary,hp2/hp3/arbitrary,hp3/hp4/arbitrary} {
          \draw [\how] (\from) -- (\to);
          \draw [mapsto] (\from) to [bend left=30] (\to);
        }
        
        \draw [mapsto,hilight] (hp1) to [bend left=30] (h);
      \end{scope}
          
      \node at (-1,0) {(b)};
    \end{scope}

      



    \begin{scope}[shift={(0,-4)}]
      \node (v1) at (0,0) [white node] {$v_1$};
      \node (v2) at (1,0) [black node] {$v_2$};
      \node (v3) at (2,0) [white node] {$v_3$};
      \node (v4) at (3,0) [black node] {$v_4$};
      \node (v5) at (4,0) [white node] {$v_5$};
      \node (v6) at (5,0) [black node] {$v_6$};
      \node (v7) at (6,0) [virtual node] {};
      
      \foreach \from/\to/\how in {v1/v2/present,v2/v3/missing,v3/v4/present,v4/v5/missing,v5/v6/present,v6/v7/missing} {
        \draw [\how] (\from) -- (\to);
        \draw [mapsto] (\from) to [bend left=30] (\to);
      }

      \draw [missing] (v2.south east) to [bend right=20] (v5.south west);
      \draw [mapsto,hilight] ([yshift=-2mm]v2.south east) to [bend right=20] ([yshift=-2mm]v5.south west);

      \node at (-1,0) {(c)};
    \end{scope}

    \begin{scope}[shift={(0,-6)}]
      \node (v1) at (0,0) [white node] {$v_1$};
      \node (v2) at (1,0) [black node] {$v_2$};
      \node (v3) at (2,0) [white node] {$v_3$};
      \node (v4) at (3,0) [black node] {$v_4$};
      \node (v5) at (4,0) [white node] {$v_5$};
      \node (v6) at (5,0) [black node] {$v_6$};
      \node (v7) at (6,0) [virtual node] {};
      
      \foreach \from/\to/\how in {v1/v2/present,v2/v3/missing,v3/v4/present,v4/v5/missing,v5/v6/present,v6/v7/missing} {
        \draw [\how] (\from) -- (\to);
        \draw [mapsto] (\from) to [bend left=30] (\to);
      }

      \draw [present] (v2.south east) to [bend right=20] (v5.south west);
      \draw [mapsto,hilight] ([yshift=-2mm]v5.south west) to [bend left=20] ([yshift=-2mm]v2.south east);

      \node at (-1,0) {(d)};
    \end{scope}

  \end{tikzpicture}
  \caption{
    Given a stable witness walk $(v_1,v_2,\dots,v_j)$ (indicated by
    blue arrows), we argue in Claim~\ref{claim:stable-set-size} that
    in certain situations we can shorten the walk via the indicated 
    red arrows, contradicting that the walk is a shortest walk. Dashed
    nodes represent arbitrarily colored nodes, dashed lines mean that
    an edge may or may not be present in the 
    graph, near-white lines represent non-edges. In~(a), we see that
    if two consecutive vertices $v_h$ and 
    $v_{h+1}$ have the same color (white in the example), then we
    could shorten the walk by returning directly from $v_{h+1}$
    to~$v_h$ (and stopping there). Hence, colors must alternate as
    shown in~(b), where we  see that if two consecutive pairs of nodes
    are connected by an edge, we can once more shorten the walk. Since
    this also hold for non-edges, edges and colors on the witness walk
    must alternate as in~(c) and~(d), assuming we start $v_1$ being
    white and with $v_1 \adj v_2$. But now (c) shows that if $v_2 \nadj v_5$, then $v_5$
    is a witness for~$v_2$, allowing us to shorten the walk, because
    the colors of and edges between $v_2$ and $v_5$ are the same as of
    and between $v_4$ and~$v_5$. On the
    other hand, in (d) we see that if $v_2\adj v_5$, then $v_2$ is a 
    witness for~$v_5$, again allowing us to return early, because
    the colors of and edges between $v_5$ and $v_2$ are the same as of
    and between $v_1$ and~$v_2$ (and also of and between $v_3$~and~$v_4$).
  }\label{figure:stable-set-size}
\end{figure}

    For the first case, assume that the color of~$v_1$ is white (for
    the case that the vertex is black, just exchange black and white
    in the following argument). Suppose $v_2$ were also white. Then
    $\phi_2$ would allow the white vertex $v_1$ to have a witness of
    the same color; but, then, $v_1$ could also serve as a witness
    for~$v_2$ (regardless of whether they are connected or not) and
    $(v_1,v_2,v_1)$ would be a stable returning $v_1$-witness walk,
    contradicting the assumption that $(v_1,\dots,v_j)$ 
    with $j > 10$ is a shortest stable $v_1$-witness walk. Thus,
    $v_2$ must be black. Repeating the argument shows 
    that $v_3$ must be white (otherwise $v_2$ would be a witness for
    $v_3$) and then $v_4$ must be black and then $v_5$ must be white
    once more.

    Again without loss of generality, assume $v_1 \adj v_2$
    (otherwise, repeat the following argument with $\adj$ and
    $\nadj$ exchanged). Then we know that the formula $\phi_2$ allows \emph{a}
    white vertex ($v_1$) to have \emph{a} black witness ($v_2$) if
    they are connected by an edge -- and since $\phi_2$ cannot
    differentiate between vertices of the same color, we get that
    \begin{align}
      \text{\emph{any white vertex~$w$ can have any black 
        vertex~$b$ as its witness whenever $w \adj
        b$.}} \label{eq-white-adj-witness}
    \end{align}

    This means, in particular, that $v_3 \nadj v_2$ since, otherwise,
    the black vertex $v_2$ could serve as a witness for the
    white~$v_3$ and $(v_1,v_2,v_3,v_2)$ would be a returning
    $v_1$-witness walk. Similarly, we also have $v_5 \nadj v_4$ for
    the same reason. In general,
    \begin{align}
      \text{\emph{any black vertex~$b$ can have
        any white vertex~$w$ as its witness whenever $w \nadj
        b$.}} \label{eq-black-nadj-witness} 
    \end{align}

    Now consider the white vertex~$v_5$ and how it is connected to the
    black vertex~$v_2$. If $v_5 \adj v_2$, then by
    \eqref{eq-white-adj-witness}, the vertex $v_2$ would be a witness
    for~$v_5$ and, thus, $(v_1,v_2,v_3,v_4,v_5,v_2)$ would be a
    $5$-vertex returning $v_1$-witness walk. If $v_5 \nadj v_2$, then
    by \eqref{eq-black-nadj-witness}, the vertex $v_5$ would be a
    witness for~$v_2$ and, thus, $(v_1,v_2,v_5,v_6,\dots,v_j)$ would
    be a \emph{shorter} returning stable $v_1$-witness walk than
    $(v_1,\dots,v_j)$. Thus, independently of whether $v_2$ and $v_5$
    are connected or not, we get a contradiction.

    For the second case, we assume that the witness walk with $j > 10$
    is returning, but to some $v_i$ with $i < 5$. We can now repeat
    all of the arguments for the first case, but starting at $v_5$
    rather than~$v_1$. For instance, the first argument is now that if
    $v_5$ is white, then $v_6$ must be black since, otherwise,
    $(v_1,\dots,v_5,v_6,v_5)$ would be a $6$-vertex returning walk
    starting at~$v_1$. By the same arguments, we also get that the
    following odd-indexed vertices must be white, while the
    even-indexed ones must be black. We can also conclude that
    \eqref{eq-white-adj-witness} and \eqref{eq-black-nadj-witness}
    must hold. Finally, we can apply the same argument as before to
    the black vertex~$v_6$ and the white vertex~$v_9$: If $v_6 \adj
    v_9$, then $v_6$ is a witness for $v_9$ and $(v_1,\dots,v_9)$ is a
    too-short $v_1$-witness walk. If $v_6 \nadj v_9$, then $v_9$ is a
    witness for $v_6$ and
    $(v_1,\dots,v_i,\dots,v_5,v_6,v_9,v_{10},\dots,v_j)$ is once more
    a shorter stable $v_1$-witness walk than $(v_1,\dots,v_j)$. Again,
    we conclude that no matter how $v_6$ and $v_9$ are connected, we
    get a contradiction.
  \end{proof}

  With the above claim, the proof of Lemma~\ref{lemma:vd-basic-eae} is complete.
\end{proof~}

Since the algorithms used to prove the next two upper bounds do not
make use of the fact that the input structure is a basic graph, we
prove them for arbitrary input structures. 

\begin{lemma}\label{lemma:vd-basic-e*a*}
  $\PVD{arb}(e^*a^*) \subseteq \Para\Class{AC}^{0}$.
\end{lemma}

\begin{proof}
  For a given formula~$\phi$ of the form $\exists x_1 \cdots \exists
  x_f \forall y_1 \cdots \forall y_g (\psi)$ for a quantifier-free
  formula~$\psi$, we show that
  $\PVD{arb}(\phi)\le^{\Para\Class{AC}^0}_{\mathrm{dtt}}\PLang{$g$-hitting-set}$,
  where the hitting set problem is defined as shown below. Since
  $\PLang{$g$-hitting-set}$ is known~\cite{BannachT20} to lie in
  $\Para\Class{AC}^{0}$, we get the claim.
  
\begin{problem}[{$\PLang{$d$-hitting-set}$ for fixed $d \in \mathbb N$}]
  \begin{parameterizedproblem}
    \instance A universe $U$ and a set $E$ of subsets $e \subseteq U$
    (called \emph{hyperedges}) with $|e| \le d$ for all $e \in E$, and
    a number~$k$.
    \parameter k.
    \question Is there a \emph{hitting set $X \subseteq V$,} meaning
    that $X \cap e \neq \emptyset$ holds for all $e \in E$, with
    $|X| \le k$?
  \end{parameterizedproblem}
\end{problem}
  
  For an arbitrary input structure $\mathcal{A}$ with universe~$A$, we
  proceed as follows: For the existentially bound variables $x_1$ to~$x_f$ we
  consider all possible assignments to them in parallel. For each of
  these, we prepare a query to the hitting set problem, resulting in
  $n^f$ queries in total. For a given assignment, which fixes each
  $x_i$ to some constant~$c_i$, replace each occurrence of $x_i$ in
  $\phi$ by~$c_i$. Build a hitting set instance $H$ as follows: The
  universe is $A \setminus\{c_1,\dots,c_f\}$. For each assignment
  $(d_1,\dots,d_g)$ of to the $g$ universally quantified variables,
  check if the formula $\psi$ is true, that is, whether $\mathcal A
  \models \psi(c_1,\dots,c_f,d_1,\dots,d_g)$. If this is not the case,
  add the hyperedge $\{d_1,\dots,d_g\} \setminus \{c_1,\dots,c_f\}$ to
  make sure that at least one element is deleted from the universe of
  $\mathcal A$ that cause this particular violation. If $\{d_1,\dots,d_g\}
  \setminus \{c_1,\dots,c_f\}$ is empty, an empty hyperedge is generated and the
  hitting set solver correctly rejects the input.

  We claim that $\mathcal{A}\in \PVD{arb}(\phi)$ iff for at least one
  of the constructed $H$ we have $(H,k) \in \PLang{$g$-hitting-set}$: For
  the first direction, let $S$ with $|S| \leq k$ be the elements of
  $\mathcal A$'s universe that we can delete, that is, for which
  $\mathcal{A}\setminus S \models \phi$. Then there are constants
  $(c_1,\dots,c_f)$ that we can assign to the existentially bound
  variables such that $\mathcal{A}\setminus S \models \forall y_1
  \cdots \forall y_g \big(\psi(c_1,\dots,c_f,y_1,\dots,y_g)\big)$. But, then,
  $S$ is a hitting set of the instance corresponding to these
  constants: If there were an edge $e \subseteq A$ with $e \cap
  S = \emptyset$ in the hitting set instance, there would be an
  assignment to the $y_i$ to elements in $A \setminus S$ that makes
  $\psi$ false, violating the assumption. 

  For the other direction, let $X$ with $|X| \leq k$ be the solution
  of one of the produced hitting set instances with $(H,k) \in
  \PLang{$g$-hitting-set}$ (at least one must exist). Then
  $\mathcal{A}\setminus X \models \phi$, since we can assign the
  existentially bound variables to the values that correspond to~$H$
  (which will not be in $X$ by construction) and there can be no
  assignment to the universally quantified variables that makes $\psi$
  false as any assignment where this would be case is hit by $X$ by
  construction and, thus, at least one element of the tuple that causes the
  violation gets removed in $\mathcal{A}\setminus X$.
\end{proof}

\begin{lemma}\label{lemma:vd-basic-e*a*e*}
  $\PVD{arb}(e^*a^*e^*) \subseteq \Para\Class{AC}^{0\uparrow}$.
\end{lemma}

\begin{proof}
  Let $\phi$ be fixed and of the form $\exists x_1 \cdots \exists x_f
  \forall y_1 \cdots \forall y_g \exists z_1 \cdots \exists z_h
  (\psi)$ for a quantifier-free formula~$\psi$. We describe a
  $\Para\Class{AC}^{0\uparrow}$-algorithm that, given an arbitrary
  input structure~$\mathcal{A}$ with universe~$A$, decides whether
  there is a set~$S$ with $|S| \leq k$ such that $\mathcal{A}\setminus
  S \models \phi$. 

  Now, we have for each assignment to the universally quantified
  variables a witness which is bound by the block of $h$ existential
  quantifiers. The problem compared to the $e^*a^*$-fragment is that
  by the deletion of elements, we could potentially destroy witnesses
  needed to satisfy other assignments.
  Because of this, we use a direct search tree algorithm to resolve
  violations of the universal quantifiers.

  In detail, we once more consider all possible assignments
  $(c_1,\dots,c_f)$ to the $x_i$ in parallel. Then we use $k$ layers
  to find and resolve 
  violations: At the start of each layer, we will already have fixed
  a set $D$ of vertices that we wish to delete, starting in the first layer with $D = \emptyset$. Then in the layer, we
  find the (for example, lexicographically) first assignment
  of the~$y_i$ to elements $(d_1,\dots,d_f)$ that all lie in $A
  \setminus D$ for which we cannot find an assignment of the $z_i$ to
  elements $(e_1,\dots,e_h)$  in $A
  \setminus D$ such that $\mathcal A \setminus D \models
  \psi(c_1,\dots,c_f,d_1,\dots,d_g,e_1,\dots,e_h)$. When we cannot find
  such an assignment, we can accept since we have found a~$D$ for
  which  $\mathcal A \setminus D \models \phi$ holds. Otherwise, we
  \emph{have to} delete one of the elements in
  $\{d_1,\dots,d_g\}\setminus \{c_1,\dots,c_f\}$ to 
  make the formula true, so we branch over these at most $g$
  possibilities, entering $g$ copies of the next layers, where the
  $i$th copy starts with $D \cup \{d_i\}$.

  Since the block of universal quantifiers has constant length, the
  number of branches in each level of the search tree is constant, so
  the total size of the search tree is at most~$g^k$. The depth of the
  search tree is bounded by the number of vertices we can delete,
  which is our parameter. In total, we get a
  $\Para\Class{AC}^{0\uparrow}$ circuit.
\end{proof}

\subparagraph*{Lower Bounds}

We now go on to show the lower bounds claimed in 
Lemma~\ref{lemma:trichotomy-basic}. The next lemmas all follow the
same rough strategy: To show that some problems that can be expressed
in the given fragments are (unconditionally) not in
$\Para\Class{AC}^{0}$, we reduce from a variant of the reachability
problem. 
%
In contrast, the last lower bound is obtained via a reduction from
$\PLang{set-cover}$, and improves a result from Fomin et
al.~\cite{FominGT20}. They establish that there is a formula
$\phi\in\Pi_3$, such that $\PVD{basic}(\phi)$ is
$\Class{W}[2]$-hard. In terms of patterns, the formula they construct
has the pattern $a^5e^{26}a$. We show that there is a formula with
pattern~$aea$ for which this holds.
 
The reachability problem that will be central for the following lower
bounds is: 

\begin{problem}[{$\PLang[]{matched-reach}$}]
  \begin{parameterizedproblem}
    \instance A directed layered graph~$G$ with vertex set
    $\{1,\dots,n\} \times \{1,\dots,k\}$, where the $i$th layer
    is $V_i \coloneq
    \{1,\dots,n\} \times \{i\}$, such that
    for each $i\in\{1,\dots,k-1\}$ the edges point to the next layer
    and they form a perfect matching between $V_i$ and $V_{i+1}$; and
    two designated vertices $s \in V_1$ and $t \in V_k$. 
    \parameter k.
    \question Is $t$ reachable from $s$ in $G$?
  \end{parameterizedproblem}
\end{problem}

(We require that in the encoding of~$G$ the vertex ``addresses''
$(i,l)$ are given explicitly as, say, pairs of binary numbers, so that
even a $\Class{AC}^0$ circuit will have no trouble determining which
vertices belong to a layer~$V_i$ or what the number~$k$ of layers is.)

Observe that the input instance can be alternatively described as a collection
of $n$ directed paths, each of length $k$. We call the paths in this graph
\emph{original paths} with \emph{original vertices and edges}. We call
the vertices in the layers $V_1$ and $V_k$ the \emph{outer vertices}
and the vertices in the layers $V_i$ for $i\in \{2, \dots, k - 1\}$
the \emph{inner vertices}. The reductions add vertices and edges to
the graphs, which will be referred to as the \emph{new vertices and
edges} (and will be indicated in yellow in figures).

\begin{fact}[\cite{BannachCT23}]\label{lemma:matched}
  $\PLang[]{matched-reach} \notin \Para\Class{AC}^0$ and, thus, for
  any problem $Q$ with $\PLang[]{matched-reach}
  \le^{\Para\Class{AC}^0}_{\mathrm{m}} Q$ we have $Q \notin  \Para\Class{AC}^0$.
\end{fact}

The proof of every lemma using a reduction from the matched reachability problem
will consist of four parts: 
\begin{enumerate}
  \item The construction of a formula~$\phi$ with the quantifier
    pattern $p$ given in the lemma.
  \item The construction of the instance for the vertex deletion
    problem $(G', k')$ from the input instance of the matched
    reachability problem $(G, s, t)$ (typically by adding new vertices
    and edges).
  \item Showing $(G, s, t) \in \PLang[]{matched-reach}$ implies $(G', k')
  \in \PVD{basic}(\phi)$, called the \emph{forward direction}.
  \item Showing $(G', k') \in \PVD{basic}(\phi)$ implies $(G, s, t) \in
  \PLang[]{matched-reach}$, called the \emph{backward direction}.
\end{enumerate}

We present the application of the above steps in detail in the
following lemma. In subsequent lemmas, which follow the same line of
arguments, but with appropriate variations in the constructions and
correctness proofs, we only highlight the differences. 

\begin{lemma}\label{lemma:vd-basic-eeae}
  $\PVD{basic}(eeae) \not\subseteq \Para\Class{AC}^{0}$.
\end{lemma}

\begin{proof}
  We want there to be a deletion
  strategy for $(G', k')$ iff in the instance $(G, s, t)$, the
  vertices $s$ and~$t$ lie on the same original path. We take~$k' =
  k$, the number of layers in~$G$, and construct a graph~$G'$ from~$G$
  by adding two special vertices~$c_1$ and~$c_2$, and regard the
  adjacency of every vertex on the original paths 
  to the vertices~$c_1$ and~$c_2$ as a $3$-coloring with colors~$i\in \{0, 1,
  2\}$. We then add appropriate gadgets at the start and the end of
  each original path, with special gadgets being added at~$s$ and
  at~$t$ (although, in this proof, their ``special gadgets'' are just
  the empty gadget). 

  \emph{The formula.} Consider the following formulas, where $\phi_{a}$
  specifies that every vertex that is neither $c_1$ nor~$c_2$ should
  be connected in a certain way to them, and $\phi_{b}$ asks that every
  vertex of color~$i$ should have a neighbor of color $(i - 1) \pmod
  3$. We encode the color~$0$ with $(x \adj c_1 \land x\nadj c_2)$,
  the color~$1$ with $(x \nadj c_1 \land x\adj c_2)$, and the
  color~$2$ with $(x \adj c_1 \land x\adj c_2)$.  

  \begin{align*}
    \phi_{a}(c_1, c_2, x) &= (c_1 \neq c_2) \land (c_1 \adj x \lor c_2\adj x)\\
    \phi_{b}(c_1, c_2, x, y) &= x\adj y \land ((x \adj c_1 \land x\adj c_2) \to (y \nadj c_1 \land y\adj c_2))\\[.25em]
    & \quad\quad\land ((x \nadj c_1 \land x\adj c_2) \to (y \adj c_1 \land y\nadj c_2))\\
    & \quad\quad\land ((x \adj c_1 \land x\nadj c_2) \to (y \adj c_1 \land y\adj c_2))\\
    \phi_{\ref{lemma:vd-basic-eeae}} &= \exists c_1 \exists c_2
    \forall x \exists y \bigl(((x \neq c_1) \land (x \neq c_2)) \to\\[.25em]
    &\qquad\qquad\qquad\qquad
    \llap{$\bigl($}(y\neq c_1) \land (y \neq c_2) \land {}\\
    &\qquad\qquad\qquad\qquad\phi_{a}(c_1, c_2, x) \land \phi_{b}(c_1, c_2, x, y)\bigr)\bigr)
  \end{align*}

\begin{figure}[htpb]
  \centering
  \begin{tikzpicture}
    \node (o_00) at (0, 0) [node] {};
    \node (o_01) at (0, 1) [node] {};
    \node (o_02) at (0, 2) [node] {};
    \node (o_03) at (0, 3) [node] {$t$};

    \node (o_10) at (1, 0) [node] {$s$};
    \node (o_11) at (1, 1) [node] {};
    \node (o_12) at (1, 2) [node] {};
    \node (o_13) at (1, 3) [node] {};

    \node (o_20) at (2, 0) [node] {};
    \node (o_21) at (2, 1) [node] {};
    \node (o_22) at (2, 2) [node] {};
    \node (o_23) at (2, 3) [node] {};

    \node (arrow) at (3, 1.5) {\Large$\mapsto$};

    \node (n_00) at (4, 0) [red node] {$0$};
    \node (n_01) at (4, 1) [green node] {$1$};
    \node (n_02) at (4, 2) [blue node] {$2$};
    \node (n_03) at (4, 3) [red node] {$t$};
    \node (nt_00) at (3.5, -1) [blue node, new] {$2$};
    \node (nt_01) at (4.5, -1) [green node, new] {$1$};

    \node (n_10) at (6, 0) [red node] {$s$};
    \node (n_11) at (6, 1) [green node] {$1$};
    \node (n_12) at (6, 2) [blue node] {$2$};
    \node (n_13) at (6, 3) [red node] {$0$};
    \node (n_14) at (6, 4) [green node, new] {$1$};

    \node (n_20) at (8, 0) [red node] {$0$};
    \node (n_21) at (8, 1) [green node] {$1$};
    \node (n_22) at (8, 2) [blue node] {$2$};
    \node (n_23) at (8, 3) [red node] {$0$};
    \node (n_24) at (8, 4) [green node, new] {$1$};
    \node (ntu_20) at (7.5, -1) [blue node, new] {$2$};
    \node (ntu_21) at (8.5, -1) [green node, new] {$1$};

    \node (u_0) at (5, 1.25) [node, new] {$c_1$};
    \node (u_1) at (5, 2.25) [node, new] {$c_2$};

    \draw[->] (o_00) -- (o_01);
    \draw[->] (o_01) -- (o_02);
    \draw[->] (o_02) -- (o_03);

    \draw[->] (o_10) -- (o_11);
    \draw[->] (o_11) -- (o_22);
    \draw[->] (o_22) -- (o_23);

    \draw[->] (o_20) -- (o_21);
    \draw[->] (o_21) -- (o_12);
    \draw[->] (o_12) -- (o_13);

    \draw (n_00) -- (n_01);
    \draw (n_01) -- (n_02);
    \draw (n_02) -- (n_03);

    \draw (n_10) -- (n_11);
    \draw (n_11) -- (n_22);
    \draw (n_22) -- (n_23);

    \draw (n_20) -- (n_21);
    \draw (n_21) -- (n_12);
    \draw (n_12) -- (n_13);

    \begin{scope}[new edge]
      \draw (nt_00) -- (nt_01);
      \draw (nt_00) -- (n_00);
      \draw (nt_01) -- (n_00);
      \draw (ntu_20) -- (ntu_21);
      \draw (ntu_20) -- (n_20);
      \draw (ntu_21) -- (n_20);
      \draw (n_13) -- (n_14);
      \draw (n_23) -- (n_24);
      \draw (u_0) -- (n_00);
      \draw (u_0) -- (n_02);
      \draw (u_0) -- (n_03);
      \draw (u_0) .. controls (5.2,-2) and (4.5, -1.5) .. (nt_00);
      \draw (u_0) -- (n_10);
      \draw (u_0) -- (n_12);
      \draw (u_0) -- (n_13);
      \draw (u_1) -- (n_01);
      \draw (u_1) -- (n_02);
      \draw (u_1) .. controls (4.1,.7) and (4.6,-.7) .. (nt_00);
      \draw (u_1) .. controls (4.5,1) .. (nt_01);
      \draw (u_1) -- (n_11);
      \draw (u_1) -- (n_12);
      \draw  (u_1) -- (n_14);
      
      \draw[dashed, shorten >=2cm] (n_20) -- (u_0);
      \draw[dashed, shorten >=2cm] (n_21) -- (u_1);
      \draw[dashed, shorten >=2cm] (n_22) -- (u_0);
      \draw[dashed, shorten >=2cm] (n_22) -- (u_1);
      \draw[dashed, shorten >=2cm] (n_23) -- (u_0);
      \draw[dashed, shorten >=2cm] (n_24) -- (u_1);
      \draw[dashed, shorten >=2cm] (ntu_20) -- (u_0);
      \draw[dashed, shorten >=2.5cm] (ntu_20) -- (u_1);
      \draw[dashed, shorten >=3cm] (ntu_21) -- (u_1);
    \end{scope}

  \end{tikzpicture}
  \caption{
    Example for the reduction from
    Lemma~\ref{lemma:vd-basic-eeae}. The input graph on the left is a
    directed layered graph with perfect matchings between consecutive 
    layers. The reduction maps it to the undirected graph shown right
    by forgetting about the direction of edges, by adding gadgets at
    the beginnings and ends of the paths (with special empty gadgets
    at $s$ and $t$), and by adding two special vertices $c_1$ and
    $c_2$ that are
    connected in three different ways to the other vertices,
    corresponding to three different colors. Newly added vertices and
    edges are indicated in yellow. Note that the indicated colors,
    numbers, and labels are not part of the output, they are only for explaining
    how the formula interprets the connection of the vertices to $c_1$
    and~$c_2$.
  }
  \label{figure:vd-basic-eeae}
\end{figure}
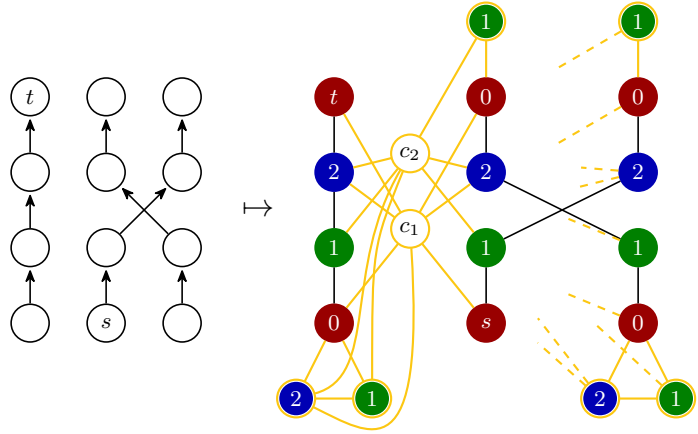

  \emph{The reduction.} On input~$(G, s, t)$ the reduction first
  checks that the graph is, indeed, a layered graph with perfect
  matchings between consecutive levels (this can easily be done by an
  $\Class{AC}^0$ circuit due to the way we encode~$G$). Then, we
  let $k'$ be the number~$k$ of layers in $G = (V,{\adj})$ and
  construct $G' = (V',\adj') $ by
  first forgetting about the direction of the edges (making the graph
  undirected). We then add the following gadgets:
  \begin{enumerate}
  \item
    At each end $v \in
    V_k$ of a path, \emph{except for $v = t$,} we add 
    a vertex~$v'$ to $V'$ and connect~$v$ to~$v'$, so $v \adj' v'$. Let 
    $V_{k+1}$ be the set of all new vertices added in this way. The
    gadget for $t \in V_k$ is empty: We do not add anything.
  \item At each
    beginning $v \in V_1$ of a path, \emph{except for $v = s$}, 
    add two vertices~$v'$ and~$v''$ to $V'$ and connect the three vertices to a
    triangle, so $v \adj' v' \adj' v'' \adj' v$. Let $V_0$ contain all
    vertices~$v'$ added in this way and let $V_{-1}$ contain all
    vertices~$v''$ added in this way. Once more, the special
    gadget for $s \in V_1$ is just the empty gadget.
  \item Finally, we add two
    further vertices~$c_1$ and~$c_2$ and connect them to the other
    vertices as follows: For $v \in V_i$ with $i\in \{-1,0,1,2, \dots, k + 1\}$:
    \begin{itemize}
    \item If $i \equiv 0 \mod 3$, let $c_1 \adj' v$.
    \item If $i \equiv 1 \mod 3$, let $c_2 \adj' v$.
    \item If $i \equiv 2 \mod 3$, let $c_1 \adj' v$ and $c_2 \adj' v$.
    \end{itemize}
  \end{enumerate}
  An example for the reduction is depicted in Figure~\ref{figure:vd-basic-eeae}. We
  claim that through this construction, the instance $(G', k')$ is in
  $\PVD{basic}(\phi_{\ref{lemma:vd-basic-eeae}})$ iff the input
  graph with vertices $s$ and $t$ is in $\PLang[]{matched-reach}$:

  \emph{Forward direction.} Suppose that~$(G, s, t) \in
  \PLang[]{matched-reach}$. We show that~$(G', k') \in
  \PVD{basic}(\phi_{\ref{lemma:vd-basic-eeae}})$: In input $G'$, just 
  delete every vertex in the original $s$-$t$-path.
  Then every vertex $v \in V_i$ for $i\in \{2, \dots, k\}$ has
  its predecessor in the original path as a neighbor, and the
  predecessor has the previous color regarding the
  ordering. Furthermore, every vertex $v\in V_1$ is part of a triangle
  where the three vertices each have a different color, so every one
  of these three vertices has a neighbor of the previous color.

  \emph{Backward direction.} Suppose that $(G', k') \in
  \PVD{basic}(\phi_{\ref{lemma:vd-basic-eeae}})$. We show that $(G, s, t) \in
  \PLang[]{matched-reach}$. By assumption, there is a set $D$ of size
  $|D| \le k = k'$ such that $G' \setminus D$ is a model of
  $\phi_{\ref{lemma:vd-basic-eeae}}$. Observe that $c_1 \notin D$ and
  $c_2 \notin D$ must hold since they are the only vertices satisfying
  the formula part~$\phi_a$, which requires that there are two
  different vertices that are connect to everyone else. On the other
  hand, we \emph{have to} delete~$s$, since by construction, it has no
  neighbor with the previous color ($s$ has color~$0$, the successor
  of~$s$ has color~$1$). But, now, the 
  successor of~$s$ has no neighbor of the previous color, so we have
  to delete it as well. We have to continue for the whole original
  path of~$s$, so $D$ has to contain at least the vertices on the
  original path starting at~$s$, which encompasses
  $k$~vertices. If the last vertex $v \in V_k$ on the original path
  starting at $s$ is not~$t$ (that is, if $t$~is not reachable
  from~$s$), then there is another vertex $v' \in V_{k+1}$ with $v
  \adj' v'$ and we also have to delete~$v'$, contradicting the
  assumption that we only have to delete $k$ vertices. Thus, $t$ must
  be reachable from~$s$.
\end{proof}

\begin{lemma~}\label{lemma:vd-basic-aae}
  $\PVD{basic}(aae) \not\subseteq \Para\Class{AC}^{0}$.
\end{lemma~}

\begin{proof~}
  We reduce from the problem $\PLang[]{matched-reach}$ to a problem
  in $\PVD{basic}(aae)$. 
  
  \emph{The formula.} Let
  \begin{align*}
    \phi_{\ref{lemma:vd-basic-aae}} \coloneq \forall x \forall y \exists z \bigl((x \adj y) \to ((x\adj z) \land (y \adj z))\bigr),
  \end{align*}
  which says that every edge should be part of a triangle.

\begin{figure}[htpb]
  \centering
  \begin{tikzpicture}
    \node (o_00) at (0, 0) [node] {};
    \node (o_01) at (0, 2) [node] {};
    \node (o_02) at (0, 4) [node] {$t$};
    \node (o_10) at (1, 0) [node] {$s$};
    \node (o_11) at (1, 2) [node] {};
    \node (o_12) at (1, 4) [node] {};
    \node (o_20) at (2, 0) [node] {};
    \node (o_21) at (2, 2) [node] {};
    \node (o_22) at (2, 4) [node] {};

    \node (arrow) at (3, 2) {\Large$\mapsto$};

    \node (n_00) at (5, 0) [node] {};
    \node (n_01) at (5, 2) [node] {};
    \node (n_02) at (5, 4) [node] {$t$};
    \node (n_10) at (8, 0) [node] {$s$};
    \node (n_11) at (8, 2) [node] {};
    \node (n_12) at (8, 4) [node] {};
    \node (n_13) at (8, 6) [node, new] {};
    \node (n_20) at (11, 0) [node] {};
    \node (n_21) at (11, 2) [node] {};
    \node (n_22) at (11, 4) [node] {};
    \node (n_23) at (11, 6) [node, new] {};

    \foreach \from/\to/\shift in {%
      n_00/n_01/-1,%
      n_01/n_02/1,%
      n_10/n_11/1,%
      n_11/n_12/-1,%
      n_12/n_13/1,%
      n_20/n_21/-1,%
      n_21/n_22/1,%
      n_22/n_23/-1%
    } {
      \node (\to_0) at ([shift={(\shift,0.25)}]\from) [small node, new] {};
      \node (\to_1) at ([shift={(\shift,0.75)}]\from) [small node, new] {};
      \node (\to_2) at ([shift={(\shift,1.25)}]\from) [small node, new] {};
      \node (\to_3) at ([shift={(\shift,1.75)}]\from) [small node, new] {};

      \draw [new edge]
        (\from) -- (\to_0) -- (\to)
        (\from) -- (\to_1) -- (\to)
        (\from) -- (\to_2) -- (\to)
        (\from) -- (\to_3) -- (\to);
    }

    \node (n_10_0) at (7, 0) [small node, new, label=left:$u^s_1$] {};
    \node (n_10_1) at (7, 0.5) [small node, new, label=left:$u^s_2$] {};
    \node (n_10_2) at (7, 1) [small node, new] {};
    \node (n_10_3) at (7, 1.5) [small node, new, label=left:$u^s_{k+1}$] {};

    \draw[->] (o_00) -- (o_01);
    \draw[->] (o_01) -- (o_02);
    \draw[->] (o_10) -- (o_11);
    \draw[->] (o_21) -- (o_12);
    \draw[->] (o_20) -- (o_21);
    \draw[->] (o_11) -- (o_22);

    \draw (n_00) -- (n_01);
    \draw (n_01) -- (n_02);
    \draw (n_10) -- (n_11);
    \draw (n_21) -- (n_12);
    \draw[new edge] (n_12) -- (n_13);
    \draw (n_20) -- (n_21);
    \draw (n_11) -- (n_22);
    \draw[new edge] (n_22) -- (n_23);

    \draw[new edge] (n_10) edge (n_10_0) edge (n_10_1) edge (n_10_2) edge (n_10_3);

  \end{tikzpicture}
  \caption{
    Example for the reduction from Lemma~\ref{lemma:vd-basic-aae},
    where, as in Figure~\ref{figure:vd-basic-eeae}, newly added
    vertices and edges are shown in yellow. The reduction adds a
    vertex at the upper end of each path except at~$t$, adds $k+1$
    ``dangling edges'' at $s$ that enforce $s$ to be deleted, and
    for each edge between some $v$ and $v'$ adds $k+1$ ``parallel
    triangles'' which enforces that if $v$ is deleted, the resulting
    $k+1$ dangling edges enforce that $v'$ is also deleted. Once more,
    any labels or colors in the figure are for illustration purposes
    only and are not part of the output.
  }
  \label{figure:vd-basic-aae}
\end{figure}

  \emph{The reduction.} Let $(G,s,t)$ and $k$ be given. As before, we
  check that the instance is valid (is layered and consecutive layers
  form perfect matchings), set $k' = k$, forget about the
  direction of the edges, and start adding gadgets.
  \begin{enumerate}
  \item
    For each end $v \in V_k$ of a path, except for~$v=t$, add a vertex
    $v' \in V_{k+1}$ and connect it to~$v$, so $v' \adj' v$. Once
    more, do not add anything to~$t$.
  \item For the $v \in V_1$ at the beginning of paths, nothing is
    done, except for $v=s$, where we add $k + 1$ new vertices $u^s_1,
    \dots, u^s_{k + 1}$ and connect them to~$s$.
  \item
    For each edge $v \adj' v'$ except for those added to~$s$, add~$k +
    1$ vertices $u^{v}_1$ to $u^v_{k+1}$ and connect them to both $v$
    and~$v'$ to form a triangle, that is, let $v \adj' u^v_i \adj' v'
    \adj' v$ be a triangle. 
  \end{enumerate}
  An example for the reduction is given in Figure~\ref{figure:vd-basic-aae}.

  \emph{Forward direction.} Suppose $(G, s, t) \in
  \PLang[]{matched-reach}$. To see that $(G', k') \in
  \PVD{basic}(\phi_{\ref{lemma:vd-basic-aae}})$, the vertex deletion
  strategy is to delete the vertices from the original path starting
  at~$s$ and ending at~$t$ (since $t$ is reachable by
  assumption). This path encompasses exactly $k$ vertices. After these
  deletions, every edge is part of a triangle: For the vertices~$v$ on
  the original path of~$s$, only the added vertices $u^v_1, \dots,
  u^v_{k + 1}$ remain, but they have degree~$0$, so no edges are left
  that need to be part of any triangles. The other original paths
  remain unmodified and every edge was already part of a triangle by
  construction. 

  \emph{Backward direction.} Suppose $(G', k') \in
  \PVD{basic}(\phi_{\ref{lemma:vd-basic-aae}})$. We show that we have $(G, s, t) \in
  \PLang[]{matched-reach}$. For $G'$ to be a model of $\phi_{\ref{lemma:vd-basic-aae}}$,
  after the deletion of $k$ vertices, we have to have deleted $s$, because otherwise,
  we would not be able to remove the $k + 1$ edges to the vertices
  $u^s_1, \dots, u^s_{k + 1}$ which are not part of a triangle. Now,
  let $v$ be the successor of $s$ in the original path. After the 
  deletion of $s$, we have $k + 1$ edges that are not part of a triangle
  between $v$ and the vertices $u^v_1, \dots, u^v_{k + 1}$, so we have to delete
  $v$ as well and so on for the whole original path of $s$. Now, if $t$ was not
  in the original path of $s$, we would have to delete $k + 1$
  vertices, a contradiction. 
\end{proof~}

\begin{lemma~}\label{lemma:vd-basic-aee}
  $\PVD{basic}(aee) \not\subseteq \Para\Class{AC}^{0}$.
\end{lemma~}

\begin{proof~}
  We again reduce from the problem $\PLang[]{matched-reach}$ to a problem
  in $\PVD{basic}(aee)$. 
  
  \emph{The formula.} Now, consider the formula
  \begin{align*}
    \phi_{\ref{lemma:vd-basic-aee}} = \forall x \exists y_1 \exists
    y_2 \bigl((x \adj y_1) \land (x\adj y_2) \land (y_1 \neq
    y_2)\bigr), 
  \end{align*}
  which requires that every vertex has degree at least~$2$.
  
\begin{figure}[htpb]
  \centering
  \begin{tikzpicture}[x=8mm]
    \node (o_00) at (0, 0) [node] {};
    \node (o_01) at (0, 1) [node] {};
    \node (o_02) at (0, 2) [node] {};
    \node (o_03) at (0, 3) [node] {$t$};
    \node (o_10) at (1, 0) [node] {$s$};
    \node (o_11) at (1, 1) [node] {};
    \node (o_12) at (1, 2) [node] {};
    \node (o_13) at (1, 3) [node] {};
    \node (o_20) at (2, 0) [node] {};
    \node (o_21) at (2, 1) [node] {};
    \node (o_22) at (2, 2) [node] {};
    \node (o_23) at (2, 3) [node] {};
    \node (o_30) at (-1, 0) [node] {};
    \node (o_31) at (-1, 1) [node] {};
    \node (o_32) at (-1, 2) [node] {};
    \node (o_33) at (-1, 3) [node] {};

    \node (arrow) at (3, 1.5) {\Large$\mapsto$};

    \node (n_00) at (5, 0) [node] {};
    \node (n_01) at (5, 1) [node] {};
    \node (n_02) at (5, 2) [node] {};
    \node (n_03) at (5, 3) [node] {$t$};

    \node (n_10) at (6, 0) [node] {$s$};
    \node (n_11) at (6, 1) [node] {};
    \node (n_12) at (6, 2) [node] {};
    \node (n_13) at (6, 3) [node] {};

    \node (n_20) at (7, 0) [node] {};
    \node (n_21) at (7, 1) [node] {};
    \node (n_22) at (7, 2) [node] {};
    \node (n_23) at (7, 3) [node] {};

    \node (n_30) at (4, 0) [node] {};
    \node (n_31) at (4, 1) [node] {};
    \node (n_32) at (4, 2) [node] {};
    \node (n_33) at (4, 3) [node] {};

    \coordinate (v_0) at (8, 1.5);

    \foreach \from/\to in {
      00/11,
      01/32,
      02/03,
      10/21,
      11/12,
      12/13,
      20/01,
      21/22,
      22/23,
      30/31,
      31/02,
      32/33%
    } {
      \draw [->] (o_\from) -- (o_\to);
      \draw (n_\from) -- (n_\to);
    }

    \foreach \from/\dir/\pos in {
      n_13/7/-0.5,
      n_23/6/-1.5,
      n_33/8/1.5,
      n_00/-7/0.5,
      n_20/-6/-1.5,
      n_30/-8/1.5
    } {
      \draw [new edge, rounded corners] (\from) -- ++(0mm,\dir mm) -| ([xshift=\pos mm]v_0);
    }

    \node at (v_0) [node, new] {$v$}; 
  \end{tikzpicture}
  \caption{
    Example for the reduction from Lemma~\ref{lemma:vd-basic-aee},
    using the same conventions as Figures \ref{figure:vd-basic-eeae}
    and~\ref{figure:vd-basic-aae}. The reduction simply adds a vertex
    that is newly connected to all beginnings and all ends of paths,
    except for $s$ and $t$. As $s$ has only a single neighbor, we
    need to delete~$s$ and then also that neighbor and then its
    neighbor and so forth. Similarly 
    for~$t$, meaning that unless $s$ and $t$ are on the same original
    path, we need to delete $2k > k$ vertices in order to ensure that
    all vertices have degree at least~$2$.
  }
  \label{figure:vd-basic-aee}
\end{figure}
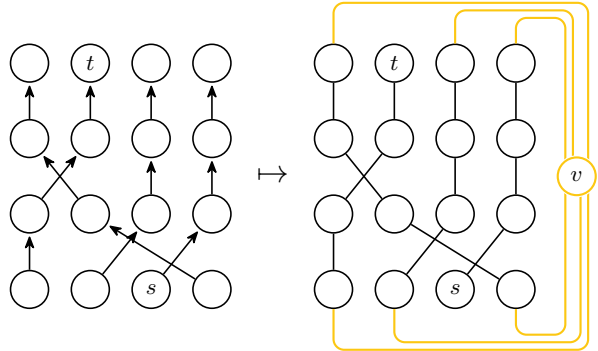

  \emph{The reduction.} For
  the reduction, we may assume without loss of generality that $k\geq 1$. Then,
  on input $(G, s, t)$, we set~$k' = k$, again forget the edge direction,
  making the graph undirected, and add a single vertex $v$, which we connect to
  every vertex from $V_1$ and $V_k$, except to $s$ and $t$. Now, every vertex
  except $s$ and $t$ has degree at least~$2$. An example for the
  reduction is given in Figure~\ref{figure:vd-basic-aee}. 
  
  \emph{Forward direction.} Assume $(G, s, t) \in
  \PLang[]{matched-reach}$. To see that $(G', k') \in
  \PVD{basic}(\phi_{\ref{lemma:vd-basic-aee}})$ holds, delete
  the $k$ vertices of the original path from $s$ to~$t$. Then every 
  vertex has degree at least~$2$: The inner vertices of the original
  graph have their predecessor and successor as neighbors, the
  vertices in $V_1$ have their successor and $v$ as neighbors, and the
  vertices in $V_k$ have their predecessor and $v$ as neighbors.

  \emph{Backward direction.} Assume $(G', k') \in
  \PVD{basic}(\phi_{\ref{lemma:vd-basic-aee}})$. Since the vertices
  $s$ and~$t$ each have degree~$1$, any deletion strategy has to
  delete them both to make the formula true. But 
  now, the successor of~$s$ and the predecessor of~$t$ have in turn
  each degree~$1$, so we have to delete them as well to make the
  formula true and so on. If $t$ was on a different original path
  as~$s$, we would have to delete at least $2k$ vertices, a
  contradiction. Thus, $(G, s, t) \in
  \PLang[]{matched-reach}$.
\end{proof~}

\begin{lemma~}\label{lemma:vd-basic-aea}
  $\PVD{basic}(aea)$ contains a  $\Class{W}[2]$-hard problem.
\end{lemma~}

\begin{proof~}
  This time, we reduce from a different problem, namely from the
  following version of the set cover problem, which is known~\cite[page 464]{DowneyF99}
  to be $\Class{W}[2]$-hard:
  \begin{problem}[{$\PLang{set-cover}$}]
  \begin{parameterizedproblem}
    \instance
      An undirected bipartite graph $G = (S \mathbin{\dot\cup} U, \adj)$
      with shores $S$ and~$U$ and a number~$k$.
    \parameter $k$
    \question Is there a cover $C \subseteq S$ with $|C| \le k$
    of~$U$, meaning that for each $u \in U$ there is an $s \in  S$
    with $u \adj s$? 
  \end{parameterizedproblem}
  \end{problem}

  \emph{The formula.}
  We use the following formula with pattern~$aea$:
  \begin{align*}
    \phi_{\ref{lemma:vd-basic-aea}} = \forall x \exists y\forall z
    \bigl((x\adj y \land (y\adj z\to x\nadj z)) \lor (x = z)\bigr) 
  \end{align*}
  This formula states that ``every vertex should have a neighbor such that
  there is no triangle of which both are part.''

\begin{figure}
  \centering
  \begin{tikzpicture}[
      right part/.style={xshift=6cm},y=2.5cm,
      node/.append style={minimum size=6mm, inner sep=0pt},
      large/.style={minimum size=7mm},
    ]
    
    \foreach \i in {1,2,3} {
      \node (s\i) at (0,\i) [node] {$s_\i$};

      \node (ns\i)    at (0,\i) [node]      [right part,shift={( 0mm,-4mm)}] {$s_\i$};
      \node (ns\i')   at (0,\i) [node, new] [right part,shift={( 0mm,4mm)}] {$s_\i'$};
      \node (ns\i'')  at (0,\i) [node, new] [right part,shift={(-8mm,4mm)}] {$s_\i''$};
      \node (ns\i''') at (0,\i) [node, new] [right part,shift={(-8mm,-4mm)}] {$s_\i'''$};

      \draw [new edge] (ns\i) -- (ns\i') -- (ns\i'') -- (ns\i''') -- (ns\i);
    }

    \foreach \i/\name in {1/u,2/v} {
      \node (\name)   at (0,\i) [green node,shift={(1,.5)}] {$\name$};

      \node (\name1)  at (0,\i) [green node, large, new]  [right part,shift={( 1,.5)}] {$\name_1$};
      \node (\name2)  at (0,\i) [green node, large, new]  [right part,shift={( 1.75,.5)}] {$\name_2$};
      \node (\name3)  at (0,\i) []  [right part,shift={( 2.5,.5)}] {$\dots$};
      \node (\name4)  at (0,\i) [green node, large, new]  [right part,shift={( 3.25,.5)}] {$\name_{k+1}$};
    }

    \foreach \s/\u/\anc in {1/u/south,2/u/north,2/v/south,3/v/north} {
      \draw (s\s) -- (\u);

      \draw [new edge]
      (ns\s) -- (\u1.\anc) -- (ns\s')
      (ns\s) -- (\u2.\anc) -- (ns\s')
      (ns\s) -- (\u4.\anc) -- (ns\s');
    }

    \node (arrow) at (3,2) {\Large$\mapsto$};
  \end{tikzpicture}
  \caption{
   Example for the reduction from Lemma~\ref{lemma:vd-basic-aea},
    using the same conventions as in the previous figures. The
    reduction gets a bipartite graph as input,
    $(\{s_1,s_2,s_3\}\cup\{u,v\},\adj)$ in the example with a white
    shore $S$ and a green shore~$U$. Each $s\in S$ is made part of a
    length-4 cycle, while for each element of $U$ exactly $k+1$ copies
    are added to the new graph. Each edge $s \adj u$ gets replaced by
    $2k+1$ edges, namely $u_i \adj' s$ and $u_u \adj' s'$ for all
    copies $u_i$ of~$u$. The size-$1$ set cover $C = \{s_2\}$
    corresponds to the fact that deleting exactly $s_2$ (or exactly
    $s_2'$) from the right graph yields a graph in which each vertex
    has an incident edge that is not part of a triangle. The same is
    true for the size-$2$ set cover $C = \{s_1,s_3\}$. In contrast,
    $C = \{s_1\}$ is not a set cover as $v$ is not covered and,
    indeed, all four incident edges of~$v_1$ (namely $s_2 \adj' v_1$,
    $s_2' \adj' v_1$, $s_3\adj' v_1$, and $s_3'\adj' v_1$) are part of
    triangles, if we  delete none of $s_2$, $s_s'$, $s_3$, or~$s_3'$.
  }
  \label{figure:vd-basic-aea}
\end{figure}
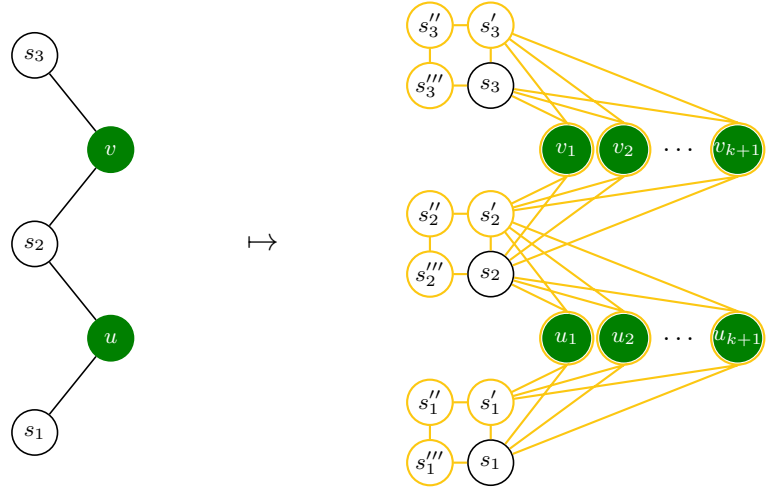

  \emph{The reduction.} Let $(S \mathbin{\dot\cup} U, \adj,k)$ be given as
  input.  The reduction outputs $k' = k$ together with the undirected
  graph $G = (V',\adj')$ constructed as follows:
  \begin{itemize}
  \item For each $s \in S$, add $s$ to $V'$ and also three more
    vertices $s',s'',s'''$ and connect them in a cycle, so $s \adj' s'
    \adj' s'' \adj'  s''' \adj' s$.
  \item For each $u \in U$, add $k+1$ copies $u_1,\dots, u_{k+1}$ of~$u$
    to~$V'$.  
  \item Whenever $u \adj s$ holds, let all $u_i$ form a triangle with $s$
    and~$s'$ in the new graph, that is, for $i\in\{1,\dots, k+1\}$ let
    $u_i \adj' s$ and $u_i \adj' s'$. 
  \end{itemize}
  An example for the reduction is shown in Figure~\ref{figure:vd-basic-aea}.

  \emph{Forward direction.}
  Let  $(S \mathbin{\dot\cup} U,\adj,k) \in \PLang{set-cover}$ be given. We need
  to show that $(G', k') \in
  \PVD{basic}(\phi_{\ref{lemma:vd-basic-aea}})$ holds. Let $C
  \subseteq S$ with $|C| \le k$ cover~$U$. We claim that $G' \setminus
  C \models \phi_{\ref{lemma:vd-basic-aea}}$, that is, removing
  all $s \in C$ from~$V'$ destroys all triangles that could violate
  $\phi_{\ref{lemma:vd-basic-aea}}$. Let us go over the different
  vertices still left in $V' \setminus C$:
  \begin{claim}
    For all $s \in S$, each vertices $s,s',s'',s'''$ has an incident
    edge in~$G'$ that is not part of a triangle. This still holds for 
    $s',s'',s'''$ in $G' \setminus C$ for $s\in C$, that is, for the
    remaining vertices of the cycles where $s$ is deleted.
  \end{claim}
  \begin{proof}
    The vertices form a cycle with four edges and of these, only
    $s \adj' s'$ is part of any triangles. Thus, the claimed edges are
    $s \adj' s'''$ for $s$ and $s' \adj' s''$ for $s'$ and $s'' \adj'
    s'''$ for $s''$ and $s''' \adj' s''$ for $s'''$. 
  \end{proof}

  \begin{claim}
    For each $u \in U$, for each of its copies $u_i$ there is an
    incident edge in $G' \setminus C$ that is not part of a triangle
    in $G' \setminus C$.
  \end{claim}
  \begin{proof}
    Since $C$ is a cover, there must be an $s \in C$ with $u \adj
    s$. But, then, $u_i \adj' s'$ by construction and this edge is no
    part of a triangle in $G'\setminus C$ (since we deleted $s \in C$ via which the only
    triangle was formed that contained this edge).
  \end{proof}
  Put together, the two claims clearly show that after removing~$C$, all
  remaining vertices have incident edges that are not part of
  triangles. 

  \emph{Backward direction.}
  Conversely, suppose that for $G' = (V', E')$ we are given a set $D
  \subseteq V'$ with $|D| \le k$ such that $G' \setminus D \models
  \phi_{\ref{lemma:vd-basic-aea}}$. 
  For each $u \in U$, consider the copies $u_i$ for $i\in
  \{1,\dots,k+1\}$. For every $s \in S$ with $s \adj u$ there is a
  triangle $u_i \adj s \adj s' \adj u_i$, but there are no other edges
  involving~$u_i$. Hence, in order to ensure that
  $\phi_{\ref{lemma:vd-basic-aea}}$ holds, we
  either have to (1) delete~$u_i$ or (2) delete exactly one of $s$
  or~$s'$ for some $s \adj u$. Since there are $k+1$ copies of~$u$, we
  cannot use option~(1) for all copies of~$u$, so \emph{for each $u
  \in U$ there must be an $s \in S$ with $s\adj u$ such that $s$ or
  $s'$ is deleted from $G'$.} However, this means that the set $C = \{s
  \in S \mid s\in D\text{ or }s' \in D\}$ is a set cover of $(S
  \mathbin{\dot\cup} U, \adj)$ and, clearly, $|C| \le |D| \le k$. 
\end{proof~}

%
%
\section{Undirected Graphs}
\label{section:undirected-graphs}

\tcsautomoveaddto{main}{\subsection{Proofs for Section~\ref{section:undirected-graphs}}}

Whether allowing self-loops has an impact on the complexity of the
problems is hard to predict: While in the setting of Fomin et
al.~\cite{FominGT20}, the same dichotomy arises for basic and
undirected graphs, in the setting of weighted 
definability considered by Bannach et al.~\cite{BannachCT23}, one
class of problems jumps from being contained in $\Para\Class{AC}^{0}$
to containing $\Para\Class{NP}$-hard problems just by allowing
self-loops. In our setting, we get an \emph{intermediate} blow-up of
the complexities by allowing self-loops: While the tractability
frontier stays the same, the frontier of fragments that are solvable
in $\Para\Class{AC}^{0}$ shifts. 

Let us now classify the complexity of vertex deletion problems on
undirected graphs. We can use some of the upper and lower bounds
established in the section before, and only consider the differences.

\begin{theorem}[{Complexity Trichotomy for $\PVD{undir}(p)$}]\label{theorem:undirected}
  Let $p \in \{a,e\}^*$ be a pattern.
  \begin{enumerate}
  \item $\PVD{undir}(p) \subseteq \Para\Class{AC}^{0}$, if $p
    \preceq ae$ or $p \preceq e^*a^*$.
  \item $\PVD{undir}(p) \subseteq \Para\Class{AC}^{0\uparrow}$
    but $\PVD{undir}(p) \not\subseteq \Para\Class{AC}^{0}$,
    if one of 
    $eae \preceq p$, $aae \preceq p$ or $aee \preceq p$ holds, but still $p
    \preceq e^*a^*e^*$ holds.
  \item $\PVD{undir}(p)$ contains a $\Class{W[2]}$-hard
    problem, if $aea\preceq p$.
  \end{enumerate}
\end{theorem}

\begin{lemma}\label{lemma:trichotomy-undirected}\hfil
  \begin{enumerate}
    \item $\PVD{undir}(ae) \subseteq \Para\Class{AC}^{0}$.
    \item $\PVD{undir}(e^*a^*) \subseteq \Para\Class{AC}^{0}$.
    \item $\PVD{undir}(e^*a^*e^*) \subseteq \Para\Class{AC}^{0\uparrow}$.
    \item $\PVD{undir}(eae)$ contains a problem not in $\Para\Class{AC}^{0}$.
    \item $\PVD{undir}(aae)$ contains a problem not in $\Para\Class{AC}^{0}$.
    \item $\PVD{undir}(aee)$ contains a problem not in $\Para\Class{AC}^{0}$.
    \item $\PVD{undir}(aea)$ contains a $\Class{W}[2]$-hard problem.
  \end{enumerate}
\end{lemma}

\begin{proof}
  Item 1 is proven below in Lemma~\ref{lemma:vd-undirected-ae}. Items 2 and 3 follow directly from Lemmas~\ref{lemma:vd-basic-e*a*}
  and~\ref{lemma:vd-basic-e*a*e*}. Item 4 is proven below in Lemma~\ref{lemma:vd-undirected-eae}, Item 5 follows from
  Lemma~\ref{lemma:vd-basic-aae}, Item 6 from Lemma~\ref{lemma:vd-basic-aee} and Item 7 from
  Lemma~\ref{lemma:vd-basic-aea}.
\end{proof}

\begin{lemma~}\label{lemma:vd-undirected-ae}
  $\PVD{undir}(ae) \subseteq \Para\Class{AC}^{0}$.
\end{lemma~}

\begin{proof~}
  A formula with the pattern $ae$ has the following form:
  \begin{align*}
    \forall x \exists y (\phi'(x,y)).
  \end{align*}
  Comparing this with \eqref{eq-phi'} from the proof of
  Lemma~\ref{lemma:vd-basic-eae}, we see that we are in a very similar
  situation as in that lemma, when $\phi'(x,y)$ is interpreted as
  $\phi_2(s,x,y)$. Of course, $\phi_2$ also talks about adjacency
  to~$s$ in the form of atoms $s \adj x$ and $s\adj y$, while $\phi'$
  also talks about self-loops in the form of atoms $x \adj x$ and
  $y\adj y$. However, it turns out that these are in one-to-one
  correspondence: In the proof, we quickly defined a two-coloring of
  the graph, where $v$ was white if $v \adj s$ held and otherwise
  black. We now call $v$ white if $v \adj v$ holds and otherwise
  black. Since these colors are the only places in the proof where the
  atoms $v \adj s$ and now $v \adj v$ are used, this replacement is
  valid.

  In a bit more detail, let us go over the proof of
  Lemma~\ref{lemma:vd-basic-eae} once more. We first defined that a
  vertex $y \in V$ is a \emph{witness for some $x \in V\setminus
  \{s\}$ relative to~$s$} if $\phi_2(s,x,y)$ held. Our new definition
  is now simply that $y \in V$ is a witness for $x \in V$ if
  $\phi'(x,y)$ holds -- the vertex $s$ no longer used or needed, just
  like the notion of something begin ``relative to~$s$''. Next, we
  defined witness walks, which could be returning, unstable, or
  $s$-terminated. Here, we simply no longer have the option of
  $s$-termination and do not need to take it into account. Thus, a
  stable witness walk is always returning. This allows us to simplify
  the algorithm as follows:

\input{undir-ae-algo.tex}

  The correctness arguments are almost all the same as in the proof of 
  Lemma~\ref{lemma:vd-basic-eae}, only we no longer need so worry
  about~$s$. Indeed, the only place in the proof where $s$ is
  mentioned once more, is when vertices $v$ are assigned the colors white
  and black depending on whether the atoms $v \adj s$ are present or
  not. While these atoms are no longer present, we now may have the
  atoms $v\adj v$ which, in the proof of
  Lemma~\ref{lemma:vd-basic-eae} always evaluated to $\mathit{false}$
  since the graphs were free of self-loops. For basic graphs this is
  no longer the case, but, fortunately, the central
  Claim~\ref{claim:stable-set-size} remains correct if in its proof we replace
  $x\adj s$ by $x\adj x$ and $y\adj s$ by $y\adj y$.
\end{proof~}

\begin{lemma~}\label{lemma:vd-undirected-eae}
  $\PVD{undir}(eae) \not\subseteq \Para\Class{AC}^{0}$.
\end{lemma~}

\begin{proof~}

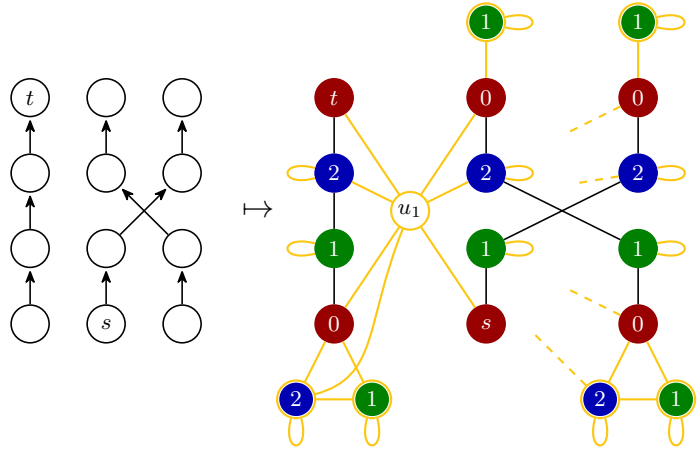
\begin{figure}[htpb]
  \centering
  \begin{tikzpicture}
    \node (o_00) at (0, 0) [node] {};
    \node (o_01) at (0, 1) [node] {};
    \node (o_02) at (0, 2) [node] {};
    \node (o_03) at (0, 3) [node] {$t$};

    \node (o_10) at (1, 0) [node] {$s$};
    \node (o_11) at (1, 1) [node] {};
    \node (o_12) at (1, 2) [node] {};
    \node (o_13) at (1, 3) [node] {};

    \node (o_20) at (2, 0) [node] {};
    \node (o_21) at (2, 1) [node] {};
    \node (o_22) at (2, 2) [node] {};
    \node (o_23) at (2, 3) [node] {};

    \node (arrow) at (3, 1.5) {\Large$\mapsto$};

    \node (n_00) at (4, 0) [red node] {$0$};
    \node (n_01) at (4, 1) [green node] {$1$};
    \node (n_02) at (4, 2) [blue node] {$2$};
    \node (n_03) at (4, 3) [red node] {$t$};
    \node (nt_00) at (3.5, -1) [blue node, new] {$2$};
    \node (nt_01) at (4.5, -1) [green node, new] {$1$};

    \node (n_10) at (6, 0) [red node] {$s$};
    \node (n_11) at (6, 1) [green node] {$1$};
    \node (n_12) at (6, 2) [blue node] {$2$};
    \node (n_13) at (6, 3) [red node] {$0$};
    \node (n_14) at (6, 4) [green node, new] {$1$};

    \node (n_20) at (8, 0) [red node] {$0$};
    \node (n_21) at (8, 1) [green node] {$1$};
    \node (n_22) at (8, 2) [blue node] {$2$};
    \node (n_23) at (8, 3) [red node] {$0$};
    \node (n_24) at (8, 4) [green node, new] {$1$};

    \node (ntu_20) at (7.5, -1) [blue node, new] {$2$};
    \node (ntu_21) at (8.5, -1) [green node, new] {$1$};

    \node (u_0) at (5, 1.5) [node, new] {$u_1$};

    \draw[->] (o_00) -- (o_01);
    \draw[->] (o_01) -- (o_02);
    \draw[->] (o_02) -- (o_03);

    \draw[->] (o_10) -- (o_11);
    \draw[->] (o_11) -- (o_22);
    \draw[->] (o_22) -- (o_23);

    \draw[->] (o_20) -- (o_21);
    \draw[->] (o_21) -- (o_12);
    \draw[->] (o_12) -- (o_13);

    \draw (n_00) -- (n_01);
    \draw (n_01) -- (n_02);
    \draw (n_02) -- (n_03);

    \draw (n_10) -- (n_11);
    \draw (n_11) -- (n_22);
    \draw (n_22) -- (n_23);

    \draw (n_20) -- (n_21);
    \draw (n_21) -- (n_12);
    \draw (n_12) -- (n_13);

    \begin{scope}[new edge]
      \draw (nt_00) -- (nt_01);
      \draw (nt_00) -- (n_00);
      \draw (nt_01) -- (n_00);
      \draw (ntu_20) -- (ntu_21);
      \draw (ntu_20) -- (n_20);
      \draw (ntu_21) -- (n_20);
      \draw (n_13) -- (n_14);
      \draw (n_23) -- (n_24);
      \draw (u_0) -- (n_00);
      \draw (u_0) -- (n_02);
      \draw (u_0) -- (n_03);
      \draw (u_0) .. controls (4.4,0) and (4.6,-.7) .. (nt_00);
      \draw (u_0) -- (n_10);
      \draw (u_0) -- (n_12);
      \draw (u_0) -- (n_13);
      
      \draw[dashed, shorten >=2cm] (n_20) -- (u_0);
      \draw[dashed, shorten >=2cm] (n_22) -- (u_0);
      \draw[dashed, shorten >=2cm] (n_23) -- (u_0);
      \draw[dashed, shorten >=2cm] (ntu_20) -- (u_0);

      \foreach \v/\w in {%
        n_01/left,%
        n_02/left,%
        n_11/right,%
        n_12/right,%
        n_14/right,
        n_21/right,%
        n_22/right,%
        n_24/right,%
        nt_00/below,%
        nt_01/below,%
        ntu_20/below,%
        ntu_21/below%
      } {
        \draw[new] (\v) to[loop \w] (\v);
      }
    \end{scope}

  \end{tikzpicture}

  \caption{
    Example for the reduction from
    Lemma~\ref{lemma:vd-undirected-eae}. The construction is nearly
    identical as the one in Figure~\ref{figure:vd-basic-eeae}: The
    only difference is that instead of adding two vertices $u_1$
    and~$u_2$ and connecting them appropriately to the other vertices
    in order to encode three colors, we only add the first
    vertex~$u_1$ (whose edges allow us to encode one bit per vertex)
    and then add self-loops to some vertices (which in combination
    with the edges to $u_1$ once more allows to encode two bits and
    hence three colors). 
  }
  \label{figure:vd-undirected-eae}
\end{figure}

  The idea for this proof is the same as in Lemma~\ref{lemma:vd-basic-eeae}, but instead of
  encoding three colors with two special vertices, we use one special vertex and
  self-loops.
  
  \emph{The formula.} Define the formula
  $\phi_{\ref{lemma:vd-undirected-eae}}$ as follows: Take the formula
  $\phi_{\ref{lemma:vd-basic-eeae}}$ from
  page~\pageref{lemma:vd-basic-eeae}, but remove the quantifier
  $\exists c_2$ (which yields the desired pattern $eae$) and replace
  each occurrence of $v \adj c_2$ by $v \adj v$ and occurrence of $v =
  c_2$ by $\mathit{false}$, where $v$ is any variable. 

  \emph{The reduction.} We only describe the difference to the
  reduction from Lemma~\ref{lemma:vd-basic-eeae}: We do not add
  $u_2$. Instead, for each vertex $v \in V'$ for which we used to have
  $v \adj' u_2$, we add a self-loop instead, so $v \adj' v$ holds
  instead. 

  \emph{Correctness.} In our construction of both the formula and of
  the graph, ``$v$ is adjacent to $c_2$'' got replaced by ``$v$ has a
  self-loop'' and, thus, the proof of Lemma~\ref{lemma:vd-basic-eeae}
  can be recycled. It only remains to argue that it is not possible
  that deleting $u_2$ would have produced solutions that are no longer 
  possible, but reviewing the proof shows that we already argued there
  that deleting $u_2$ is not possible (and neither is deleting~$u_1$).
\end{proof~}

%
%
\section{Directed~Graphs~and~Arbitrary~Structures}
\label{section:directed-graphs}

\tcsautomoveaddto{main}{\subsection{Proofs for Section~\ref{section:directed-graphs}}}

The final class of logical structures we investigate in this paper are directed
graphs. Interestingly, from the viewpoint of quantifier patterns, this class of
structures is as complex as arbitrary logical structures.

\begin{theorem}[{Complexity Trichotomy for $\PVD{dir}(p)$}]\label{theorem:directed}
  Let $p \in \{a,e\}^*$ be a pattern.
  \begin{enumerate}
  \item $\PVD{dir}(p) \subseteq \Para\Class{AC}^{0}$, if $p \preceq e^*a^*$.
  \item $\PVD{dir}(p) \subseteq \Para\Class{AC}^{0\uparrow}$
    but $\PVD{dir}(p) \not\subseteq \Para\Class{AC}^{0}$, if
    $ae \preceq p \preceq e^*a^*e^*$.
  \item $\PVD{dir}(p)$ contains a $\Class{W[2]}$-hard
    problem, if $aea\preceq p$.
  \end{enumerate}
\end{theorem}

\begin{lemma}\label{lemma:trichotomy-directed}\hfil
  \begin{enumerate}
    \item $\PVD{dir}(e^*a^*) \subseteq \Para\Class{AC}^{0}$.
    \item $\PVD{dir}(e^*a^*e^*) \subseteq \Para\Class{AC}^{0\uparrow}$.
    \item $\PVD{dir}(ae)$ contains a problem not in $\Para\Class{AC}^{0}$.
    \item $\PVD{dir}(aea)$ contains a $\Class{W}[2]$-hard problem.
  \end{enumerate}
\end{lemma}

\begin{proof}
  Items 1 and 2 follow directly from Lemmas~\ref{lemma:vd-basic-e*a*}
  and~\ref{lemma:vd-basic-e*a*e*}. Item 3 is shown in Lemma~\ref{lemma:vd-directed-ae}, and Item 4 follows from
  Lemma~\ref{lemma:vd-basic-aea}.
\end{proof}

\begin{lemma~}\label{lemma:vd-directed-ae}
  $\PVD{dir}(ae) \not\subseteq \Para\Class{AC}^{0}$.
\end{lemma~}

\begin{proof~}
  
  \emph{The formula.} Consider the formula
  \begin{align*}
    \phi_{\ref{lemma:vd-directed-ae}} \coloneq \forall x \exists y \bigl(x\adj y\bigr),
  \end{align*}
  which states that every vertex has a successor.
  
\begin{figure}
  \centering
  \begin{tikzpicture}
    \node (o_00) at (0, 0) [node] {};
    \node (o_01) at (0, 1) [node] {};
    \node (o_02) at (0, 2) [node] {};
    \node (o_03) at (0, 3) [node] {$t$};
    \node (o_10) at (1, 0) [node] {$s$};
    \node (o_11) at (1, 1) [node] {};
    \node (o_12) at (1, 2) [node] {};
    \node (o_13) at (1, 3) [node] {};
    \node (o_20) at (2, 0) [node] {};
    \node (o_21) at (2, 1) [node] {};
    \node (o_22) at (2, 2) [node] {};
    \node (o_23) at (2, 3) [node] {};

    \node (arrow) at (3, 1.5) {\Large$\mapsto$};

    \node (n_0s) at (4, -1) [node, new] {};
    \node (n_00) at (4, 0) [node] {};
    \node (n_01) at (4, 1) [node] {};
    \node (n_02) at (4, 2) [node] {};
    \node (n_03) at (4, 3) [node] {$t$};

    \node (n_10) at (5, 0) [node] {$s$};
    \node (n_11) at (5, 1) [node] {};
    \node (n_12) at (5, 2) [node] {};
    \node (n_13) at (5, 3) [node] {};

    \node (n_2s) at (6, -1) [node, new] {};
    \node (n_20) at (6, 0) [node] {};
    \node (n_21) at (6, 1) [node] {};
    \node (n_22) at (6, 2) [node] {};
    \node (n_23) at (6, 3) [node] {};

    \draw[->] (o_00) -- (o_01);
    \draw[->] (o_01) -- (o_12);
    \draw[->] (o_02) -- (o_03);

    \draw[->] (o_10) -- (o_11);
    \draw[->] (o_11) -- (o_22);
    \draw[->] (o_12) -- (o_13);

    \draw[->] (o_20) -- (o_21);
    \draw[->] (o_21) -- (o_02);
    \draw[->] (o_22) -- (o_23);

    \draw[->, new edge] (n_0s) -- (n_00);
    \draw[->] (n_00) -- (n_01);
    \draw[->] (n_01) -- (n_12);
    \draw[->] (n_02) -- (n_03);

    \draw[->] (n_10) -- (n_11);
    \draw[->] (n_11) -- (n_22);
    \draw[->] (n_12) -- (n_13);

    \draw[->, new edge] (n_2s) -- (n_20);
    \draw[->] (n_20) -- (n_21);
    \draw[->] (n_21) -- (n_02);
    \draw[->] (n_22) -- (n_23);
    \draw[->, new edge] (n_23) to[loop above] (n_23);
    \draw[->, new edge] (n_13) to[loop above] (n_13);
  \end{tikzpicture}
  \caption{
    Example for the reduction from Lemma~\ref{lemma:vd-directed-ae},
    once more using the conventions from the previous figures. The
    construction is quite simple: Add a self-loop at the end of all
    paths, except at the path ending at~$t$, and elongate all paths
    by~$1$ except for the path containing~$s$.  Since our formula
    requires that every vertex has a successor and~$t$ does not, we
    are forced to delete~$t$ and then also its predecessor and so
    on. If $t$ is reachable from~$s$, we will be done after $k$
    deletions, otherwise not.
  }
  \label{figure:vd-directed-ae}
\end{figure}
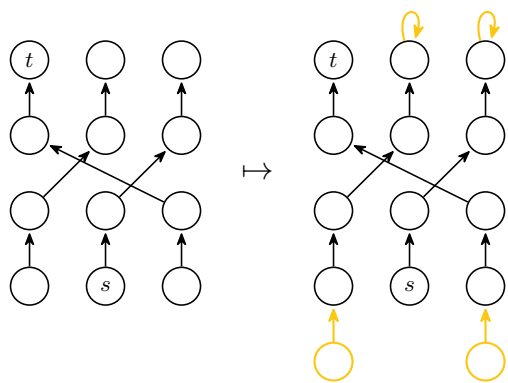

  \emph{The reduction.} On input $(G, s, t)$ we reduce
  as follows: We set~$k' = k$, and for each vertex $v\in V_k$, except for~$t$, we add a self-loop. We
  then add for every vertex $u\in V_1$, except for~$s$, a vertex~$u'$ and an edge
  $(u', u)$. An example for the reduction is given in Figure~\ref{figure:vd-directed-ae}.
  
  \emph{Forward direction.} We have that $(G, s, t) \in
  \PLang[]{matched-reach}$, and show that we have $(G', k') \in
  \PVD{dir}(\phi_{\ref{lemma:vd-directed-ae}})$. To make the
  formula true, we simply delete every vertex in the original path of~$t$, and
  since $s$ is in the same original path, we have to delete exactly $k$
  vertices. Now every vertex has a successor: The vertices in the layers $V_i$
  for $i\in \{1, \dots, k - 1\}$ have their original successor, and every vertex
  in $V_k$ has itself as a successor via the self-loop.
  
  \emph{Backward direction.} We have that $(G', k') \in
  \PVD{dir}(\phi_{\ref{lemma:vd-directed-ae}})$ and show that we
  have $(G, s, t) \in
  \PLang[]{matched-reach}$. To make the formula true, we have to delete $t$, since it has no
  successor. But then, we have also delete the predecessor of $t$, since it now
  has no successor as well, and so on. So, we have to delete all the
  vertices on the same original path as $t$. Now, if this original path did not
  begin with $s$, we would have to delete $k + 1$ vertices, a contradiction.
\end{proof~}

\section{Conclusion}
In this paper, we fully classified the parameterized complexity of vertex
deletion problems where the target property is expressible by
first-order formulas and where the inputs are basic graphs,
undirected graphs, directed graphs, or arbitrary logical structures. The
classification is based on the quantifier patterns of the formulas, and sheds
additional light on the complexity properties that emerge from these patterns:
We have seen that while the tractability barrier is the same for all logical
structures, $\PLangVD{basic}(e^*a^*e^*), \PLangVD{undir}(e^*a^*e^*),
\PLangVD{dir}(e^*a^*e^*)$ and $\PLangVD{arb}(e^*a^*e^*)$ all being tractable and
$\PLangVD{basic}(aea), \PLangVD{undir}(aea), \PLangVD{dir}(aea)$ as well as
$\PLangVD{arb}(aea)$ all containing intractable problems, in the tractable
cases, basic, undirected and directed graphs have provably different
complexities, the latter coinciding with arbitrary structures.

The granularity we gained with the viewpoint of quantifier patterns could be
useful to examine the complexity of vertex deletions problems where the property
is given by a formula of a more expressive logic: For both \emph{monadic
second-order logic} ($\Lang{mso}$) and \emph{existential second-order logic}
($\Lang{eso}$), even the model checking problem becomes $\Class{NP}$-hard.
 This would allow us to
express many more natural problems such as feedback vertex set, that have no
obvious formalization as a vertex deletion problem to plain
$\Lang{fo}$-properties. Similarly, we could allow extensions such as
transitive closure or fixed point operators. 

Compared to previous work on weighted definability, where the
objective is to instantiate a free set variable with \emph{at most,} 
\emph{exactly,} or \emph{at least} $k$~elements such that a formula
holds, we only considered deleting \emph{at most}~$k$ elements. How
does the complexity of vertex deletion problems change, if we
\emph{have to} delete exactly $k$ elements -- or, for that matter,
\emph{at least} $k$~elements?


\bibliography{main}

\tcsautomoveinsert{main}

\end{document}

%% file: basic-eae-algo.tex
\begin{lstlisting}[style=pseudocode]
input $G = (V,E)$

for $s \in V$ do $\label{line-branch}$
  $D \gets \emptyset$ $\label{line-d-start}$
  for $v \in V \setminus \{s\}$ do
    if there is no stable $v\text-$witness walk of length at most $10$ relative to $s$ then $\label{line-stable}$
      $D \gets D \cup \{v\}$ $\label{line-add}\label{line-d-end}$

  if $|D| \le k$ then $\label{line-size-check}$
    $G' \gets G \setminus D$
    if $G' \models \exists y(\phi_1(s, s, y))$ then $\label{line-final-check}$
      output ``$(G,k) \in \PLangVD{basic}(\phi)$'' and stop $\label{line-accept}$

output ``$(G,k) \notin \PLangVD{basic}(\phi)$''  $\label{line-reject}$
\end{lstlisting}

%% file: undir-ae-algo.tex
\begin{lstlisting}[style=pseudocode]
input $G = (V,E)$

$D \gets \emptyset$
for $v \in V$ do
  if there is no stable $v\text-$witness walk of length at most $10$ then
    $D \gets D \cup \{v\}$

if $|D| \le k$ then output ``$(G,k) \in \PLangVD{undir}(\phi)$'' else output ``$(G,k) \notin \PLangVD{undir}(\phi)$''
\end{lstlisting}